\documentclass{aastex63}
\usepackage{graphicx}
\usepackage{subfigure}
\usepackage{booktabs}
\usepackage{lineno}
\usepackage{longtable}

\shorttitle{huang article}
\shortauthors{huang et al.}

\begin{document}

\flushleft{Accepted by ApJ}

\title{A statistical study of the plasma and composition distribution inside magnetic clouds: 1998 - 2011}
\correspondingauthor{Jin Huang}
\email{huang$\_$heliophysics@163.com}

\author[0000-0003-0099-0406]{Jin Huang}
\affiliation{Yunnan Observatories, Chinese Academy of Sciences, Kunming 650011, People's Republic of China}
\affiliation{Shandong Provincial Key Laboratory of Optical Astronomy and Solar-Terrestrial Environment, Shandong University, Weihai 264209, People¡¯s Republic of China}
\affiliation{Institute of Space Physics, Luoyang Normal University, Luoyang 471934, People's Republic of China}
\affiliation{College of Physics and Electric Information, Luoyang Normal University, Luoyang 471934, People's Republic of China}
\affiliation{Radio Cosmology Lab, Department of Physics, Faculty of Science, University of Malaya, 50603 Kuala Lumpur, Malaysia}
\affiliation{University of Chinese Academy of Sciences, Beijing 100049, People's Republic of China}

\author{Yu Liu}
\affiliation{Yunnan Observatories, Chinese Academy of Sciences, Kunming 650011, People's Republic of China}
\affiliation{Shandong Provincial Key Laboratory of Optical Astronomy and Solar-Terrestrial Environment, Shandong University, Weihai 264209, People¡¯s Republic of China}
\affiliation{University of Chinese Academy of Sciences, Beijing 100049, People's Republic of China}

\author{Hengqiang Feng}
\affiliation{Institute of Space Physics, Luoyang Normal University, Luoyang 471934, People's Republic of China}

\author{Ake Zhao}
\affiliation{College of Physics and Electric Information, Luoyang Normal University, Luoyang 471934, People's Republic of China}

\author{Z.Z.Abidin}
\affiliation{Radio Cosmology Lab, Department of Physics, Faculty of Science, University of Malaya, 50603 Kuala Lumpur, Malaysia}

\author{Yuandeng Shen}
\affiliation{Yunnan Observatories, Chinese Academy of Sciences, Kunming 650011, People's Republic of China}

\author{Oloketuyi Jacob}
\affiliation{Yunnan Observatories, Chinese Academy of Sciences, Kunming 650011, People's Republic of China}
\affiliation{University of Chinese Academy of Sciences, Beijing 100049, People's Republic of China}

\nocollaboration{7}
\begin{abstract}

A comprehensive analysis of plasma and composition characteristics inside magnetic clouds (MCs) observed by the Advanced Composition Explorer (ACE) spacecraft from 1998 February  to 2011 August is presented. The results show that MCs have specific interior structures, and MCs of different speeds show differences in composition and structure. Compared with the slow MCs, fast MCs have enhanced mean charge states of iron, oxygen, silicon, magnesium, $\mathrm{O^{7+}/O^{6+}}$, $\mathrm{C^{6+}/C^{5+}}$, $\mathrm{C^{6+}/C^{4+}}$ and $\mathrm{Fe^{\geq16+}/Fe_{total}}$ values. For ionic species in fast MCs, a higher atomic number represents a greater enhancement of mean charge state than slow MCs. We also find that both the fast and slow MCs display bimodal structure distribution in the mean iron charge state ($\mathrm{\langle Q\rangle Fe}$), which suggests that the existence of flux rope prior to the eruption is common. Furthermore, the $\mathrm{\langle Q\rangle Fe} $, $\mathrm{Fe^{\geq16+}/Fe_{total}}$, and $\mathrm{O^{7+}/O^{6+}}$  ratio distribution inside fast MCs have the feature that the posterior peak is higher than the anterior one. This result agrees with the ``standard model" for CME/flares, by which magnetic reconnection occurs beneath the flux rope, thereby ionizing the ions of the posterior part of flux rope sufficiently by high-energy electron collisions or by direct heating in the reconnection region.

\end{abstract}


\section{Introduction} \label{sec:intro}

Coronal mass ejections (CMEs), which are the most severe explosive phenomena in the heliosphere, observed with coronagraphs \citep[e.g.,][]{2011JASTP..73.1242H, 2011JASTP..73.1187L}, play a central role in the influence of the active Sun on significant interplanetary disturbances, and geospace environment variations via the solar wind (SW).

Many CMEs, for some instances, correspond to a magnetically organized geometry of flux rope (termed as flux-rope CMEs) \citep{1996JGR...10127499C,1998ApJ...493..460G}, the occurrence of magnetic flux ropes in interplanetary space, often referred to as magnetic clouds (MCs) \citep{1981JGR....86.6673B}, which have been associated with prominence eruptions at the Sun \citep{1982GeoRL...9.1317B}. MCs are large scale, organized magnetic flux rope structures, characterized by several signatures, as reviewed for example by \citet{1990GMS....58..343G}, \citet{1997GMS....99..245N}, and \citet{2006SSRv..123...31Z}. We are now able to use a combination of the magnetic field, plasma, compositional, and energetic particle signatures to identify MCs. In order to provide a credible link between MCs and eruptive prominences of the Sun, in-depth exploration of the magnetic field structure and topology of MCs has been performed \citep[e.g.,][]{1990JGR....9511957L,1998AnGeo..16....1B,2002JGRA..107.1142H}, and their internal plasma and composition statistical studies have been conducted by some authors as well \citep[e.g.,][]{2004JGRA..109.1108R,2013SoPh..284...77K,2018SoPh..293..122O}. Most of them explored the MCs internal structure by case event studies or statistical analysis by time series.

\cite{2003JGRA..108.1239L}, hereafter Lynch03, carried out a superposed epoch analysis to explore the internal structure of the MCs. They constructed diameter cuts through MC profiles (1998 - 2001) to obtain plasma and composition quantity on average, and associate the in-situ measured quantity with a radial distance in MC cylinder model. It is a good supplement to the traditional analysis of MCs. In this study, we extend the statistical period to span a solar cycle and apply this method for more desired plasma, elemental composition, especially ionic charge state quantity to seek the statistical properties of internal MCs.

The ionic charge states of plasma in a CME are an imprint of the electron temperature distribution on a few solar radii ($R_{\odot}$) above the solar surface, and the appearance of high ionic charge states usually implies high electron temperatures. During CME eruption, magnetic reconnection occurs along the current sheet that connects the CME flux rope structure to flare loops \citep{2006ApJ...638.1110B,2013AIPC.1539..207K,2015ApJ...803...96S,2015ApJ...808L..15S}, and reconnection heats the plasma, as this process continues, the heated plasma injects into the flux rope along magnetic field lines \citep{2013AIPC.1539..207K}. On CME transits' way to the Earth, the coronal electron density continually decreases, solar wind ion expansion time is smaller than the ionization and recombination time-scales at a certain altitude. Thereafter, heavy-ion (atomic number $> 2$) charge states and elemental composition are not apparently changed, which often happens within 5 $R_{\odot}$, namely, freeze-in. Therefore, such properties can be used to infer the formation time of magnetic flux rope in MC and the current sheet temperature information during CME eruption \citep[e.g.,][]{1999JGR...10417005K,2011ApJ...740..112L,2012ApJ...760..105L}. A high/normal ionic charge state indicates high/normal current sheet temperature \citep[e.g.,][]{2013ApJ...766...65C}. In general, they provide not only one of the best tools to identify CME material in interplanetary space, but also an important way in tracing back the solar environment of the MCs origin \citep{1995Sci...268.1033G}.

There are two theories of the origin of flux-rope CMEs: (1) the flux ropes exist prior to the eruption and erupt via some mechanism; and (2) the flux rope structure was formed during the eruption. Some observations \citep[e.g.,][]{2012NatCo...3..747Z,2013ApJ...763...43C,2014ApJ...789...93C,2003ApJ...593L.137L,2013ApJ...764..125P,2014ApJ...784...48S} and numerical simulations \citep{2000JGR...105.2375L} support the former, whereas other simulations \citep{1994ApJ...430..898M} and observations \citep{2015ApJ...815...72O} hold that flux ropes can also be formed during the eruption. \citet{2016ApJS..224...27S} made a comprehensive survey of the mean iron charge state $\mathrm{\langle Q\rangle Fe}$ distributions inside MCs for solar cycle 23. Results showed that 11 in 92 MCs exhibited a bimodal distribution with both peaks higher than 12+ for $\mathrm{\langle Q\rangle Fe}$. During eruption, a pre-existing flux rope with relatively low temperature was surrounded by reconnecting magnetic field lines and heated plasma. When the corresponding MC was detected near the Earth, high-ionization-state shell and low-ionization-state center would be found \citep{2004ApJ...602..422L,2013AIPC.1539..207K,2015ApJ...804L..38S}. Such a bimodal profile of $\mathrm{\langle Q\rangle Fe}$ suggests that the flux rope in MCs existed before the CMEs' eruption. It is necessary to test the scenario in a preliminary and statistical approach with in-situ observations.

Another controversial issue associated with CMEs is which factors determine the CME velocity. \citet{2002AAS...200.3704Y}, and \citet{2005A&A...435.1149V} revealed that there is only a weak correlation between the CME apparent velocity and the peak flux of the associated flares. A different conclusion drawn by \cite{2009ApJ...705..914R} is that energetic source regions produce fast CMEs that are accompanied by larger flares whereas less energetic sources produce slow CMEs accompanied by smaller flares. This conclusion complies with the view that CME velocity is related to magnetic reconnection flux \citep{2005ApJ...634L.121Q} or magnetic field in the filament channel\citep{2006A&A...456.1153C}. In general, if CME is faster (slower) than the ambient SW, and it is decelerated (accelerated) by the SW \citep{1996JGR...10127499C}, but fast/slow MCs still correspond to fast/slow flux-rope CMEs at 1 AU \citep{1999JGR...10412515L}. Therefore, in this study, we try to investigate the plasma, elemental and charge state composition differences between the fast and the slow MCs, to improve understanding on this issue.

This study is focused on the internal structure of the plasma and ionic charge states obtained from the inherent geometry of the linear force-free magnetic field model, shedding more light on characteristic differences of fast/slow MCs by making a comprehensive survey of 124 MCs for 1998 - 2011 using Advanced Composition Explorer (ACE) \citep{1998SSRv...86..257C} spacecraft data. The study is structured as follows: section \ref{sec:data} describes data and event selection procedure, and fit the events with linear force-free model. In section \ref{sec:method}, we describe the method of inferring spatial position. Section \ref{sec:results} presents the statistical results, followed by explanations of the results and conclusion in Section \ref{sec:conclusion}.

\section{ Data description, events selection, and model fitting} \label{sec:data}

We have identified and modeled 124 MC events observed by the ACE spacecraft between 1998 February and 2011 August, when ACE/SWICS data are available, covering a solar cycle. The magnetic field magnitude data were provided by ACE/MAG every 4-mins. Helium-to-proton density ratio ($\mathrm{He/P}$), SW bulk speed, and proton temperature data were provided with a cadence of 1-hr by ACE/SWEPAM. The rest of the data came from ACE/SWICS, in which $\mathrm{O^{7+}/O^{6+}}$, $\mathrm{C^{6+}/C^{5+}}$, $\mathrm{C^{6+}/C^{4+}}$, $\mathrm{Fe/O}$, charge state of C, O, Mg, Si, Fe used 1-hr cadence data, Ne/O, Mg/O, Si/O, C/O, He/O used 2-hrs cadence data, and proton number density used 12-mins cadence data. Descriptions of related instruments were given by \citet{1998SSRv...86..613S}, \citet{1998SSRv...86..563M} and \citet{1998SSRv...86..497G}.

Our MC events come partly from ACE observations published in the KASI online MC list (see the cylinder model events in http://sos.kasi.re.kr/mc/). In addition, considering the scale of the MCs, there is little difference between WIND and ACE observations in most cases, thus we also tried to employ WIND observations as supplements. The WIND MC lists of \cite{2006AnGeo..24..215L,2011SoPh..274..345L} and \cite{2016JGRA..121.9316W} therefore were referenced. All of the candidates were checked with ACE data by visual inspection according to the criterion given by \cite{1981JGR....86.6673B}. Roughly enhanced magnetic field strength and relatively smooth change in field direction are required. Subsequently, they were fitted by static, constant-$\alpha$, cylindrically symmetric, force-free MC model \citep{1988JGR....93.7217B,1990JGR....9511957L}, which is still one of the most commonly used techniques to analyze MC to date. The magnetic force-free field with constant-$\alpha$ satisfies Eq.(1).

\begin{equation}\label{eq1}
  \nabla\times\mathbf{B}=\alpha\mathbf{B}
\end{equation}

\begin{figure}[t]
\centering
  \begin{tabular}{@{}cccc@{}}
    \includegraphics[width=.45\textwidth]{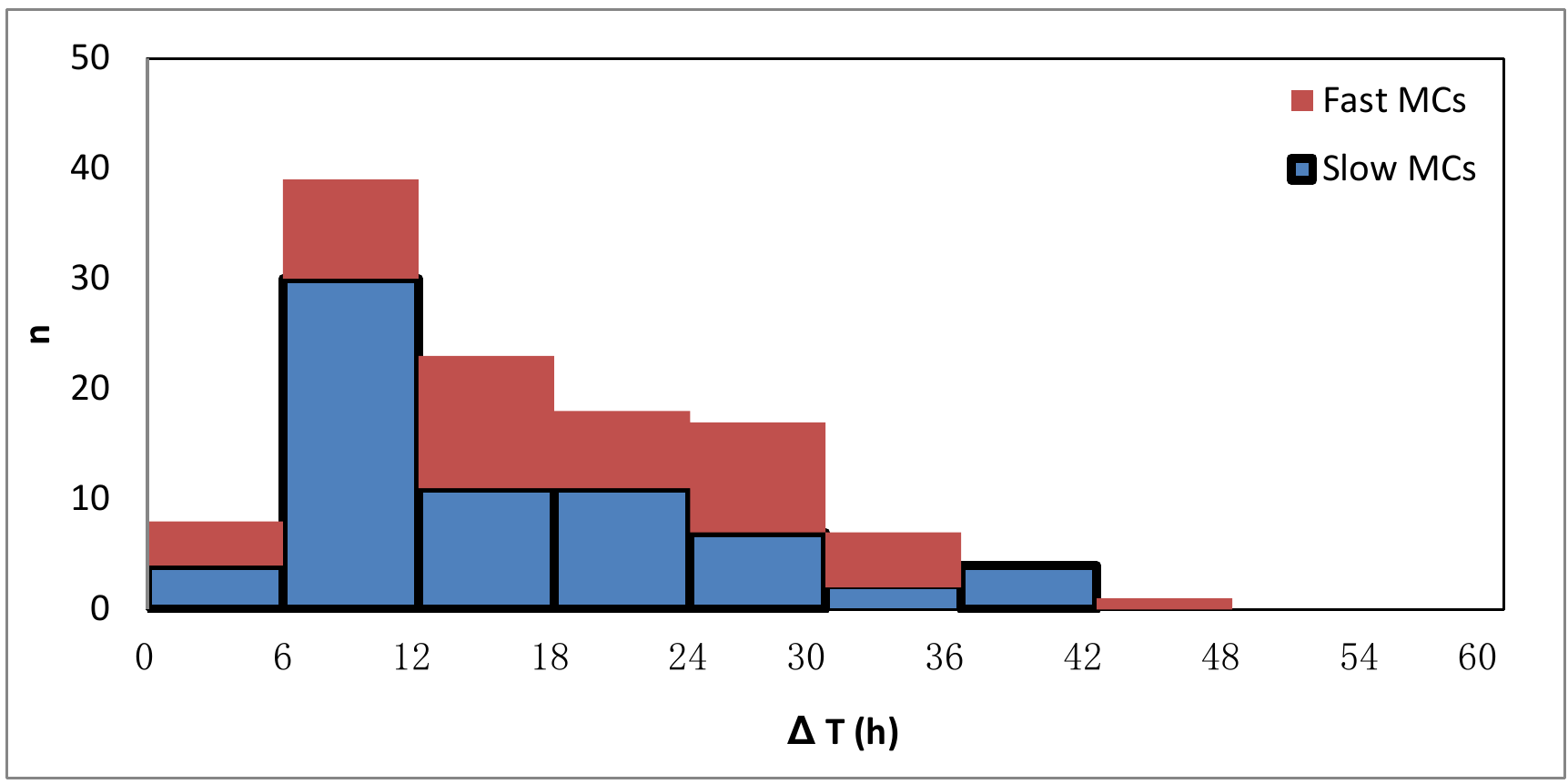} &
    \includegraphics[width=.45\textwidth]{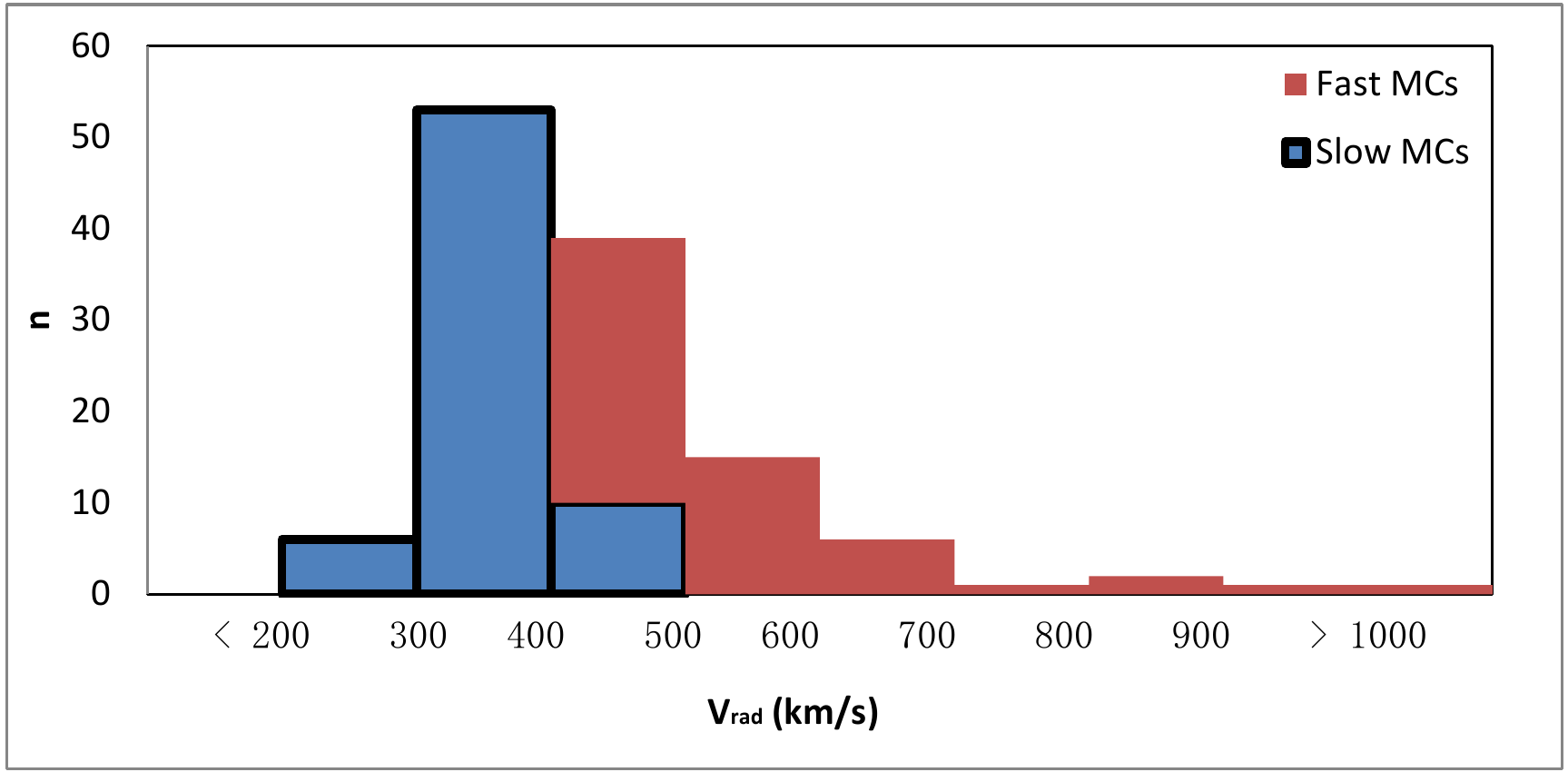} \\
    \includegraphics[width=.45\textwidth]{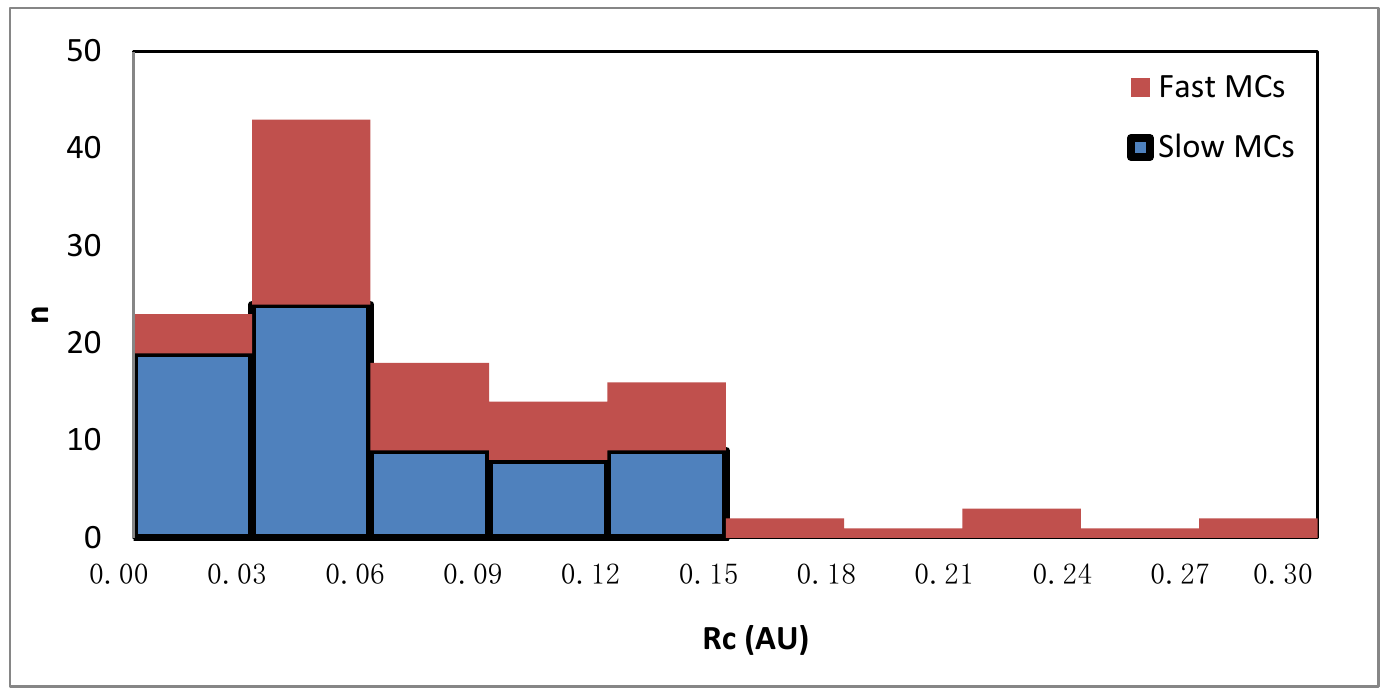} &
    \includegraphics[width=.45\textwidth]{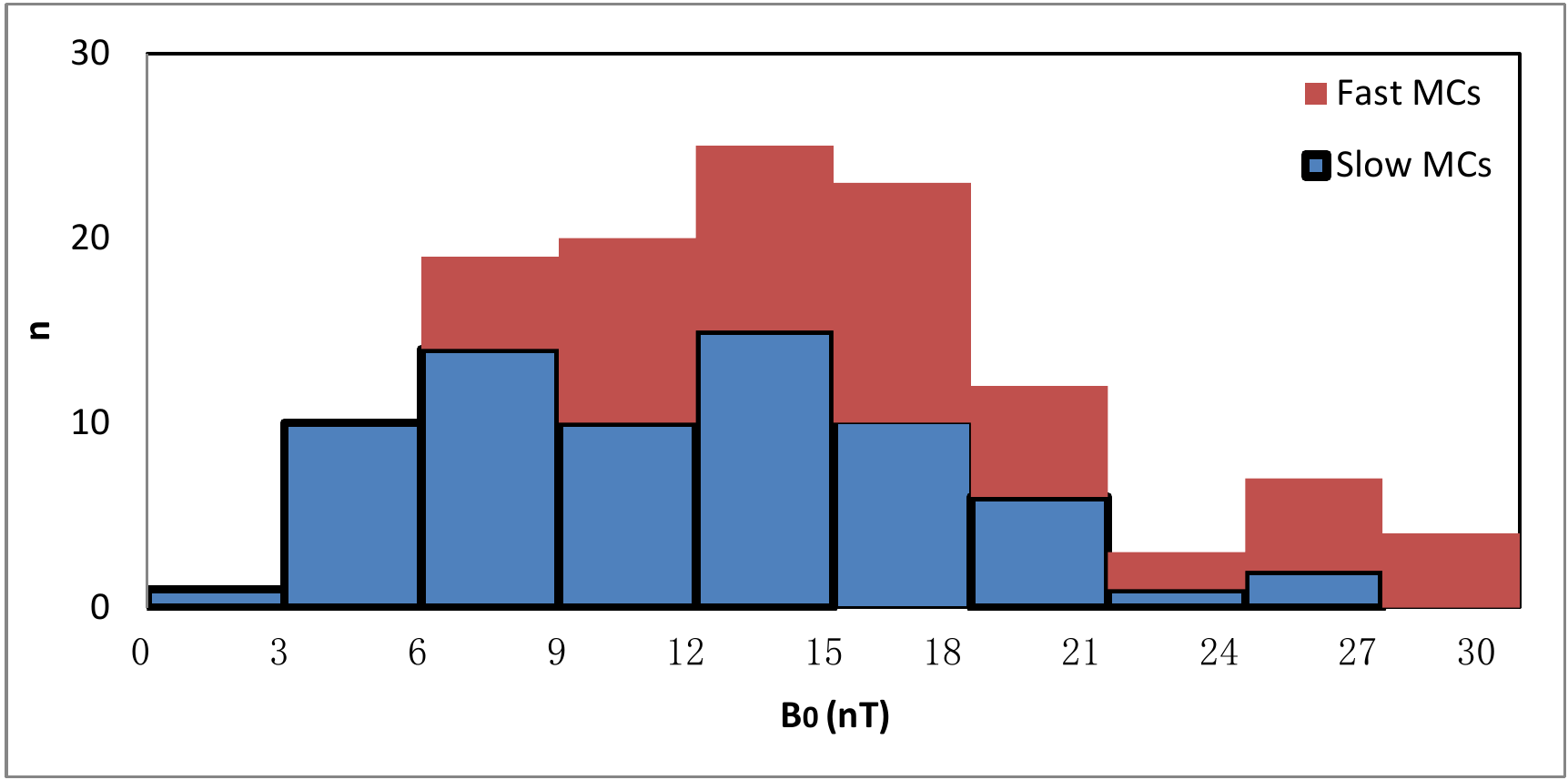}  \\
    \includegraphics[width=.45\textwidth]{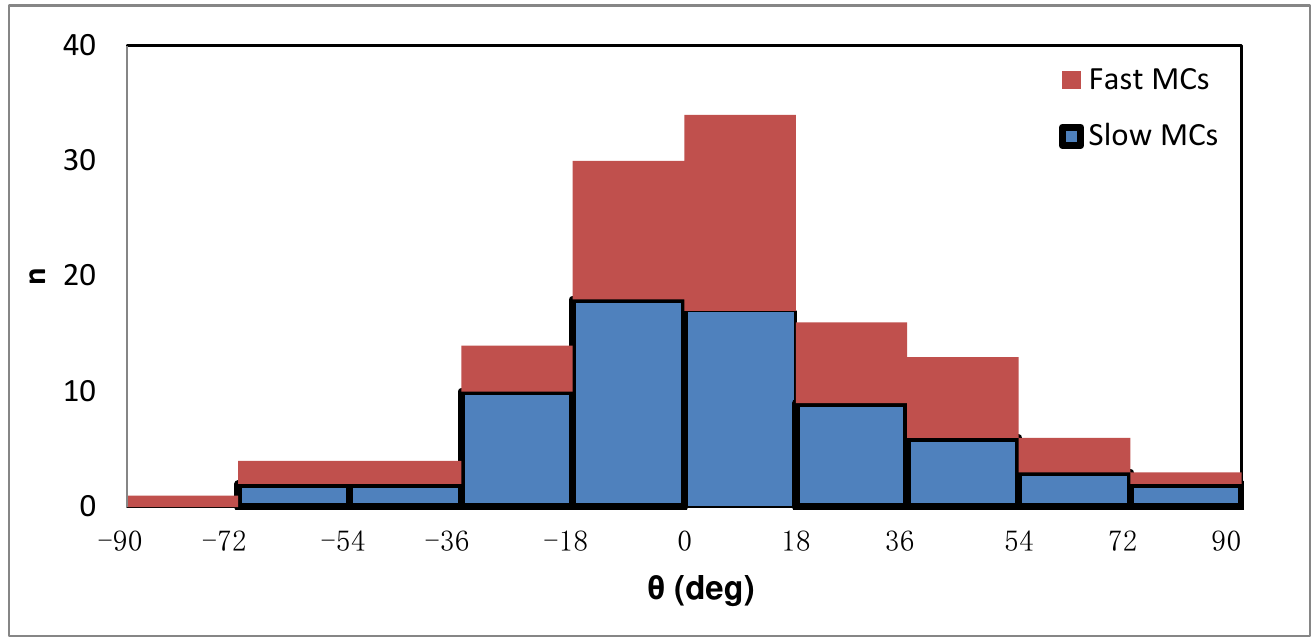} &
    \includegraphics[width=.45\textwidth]{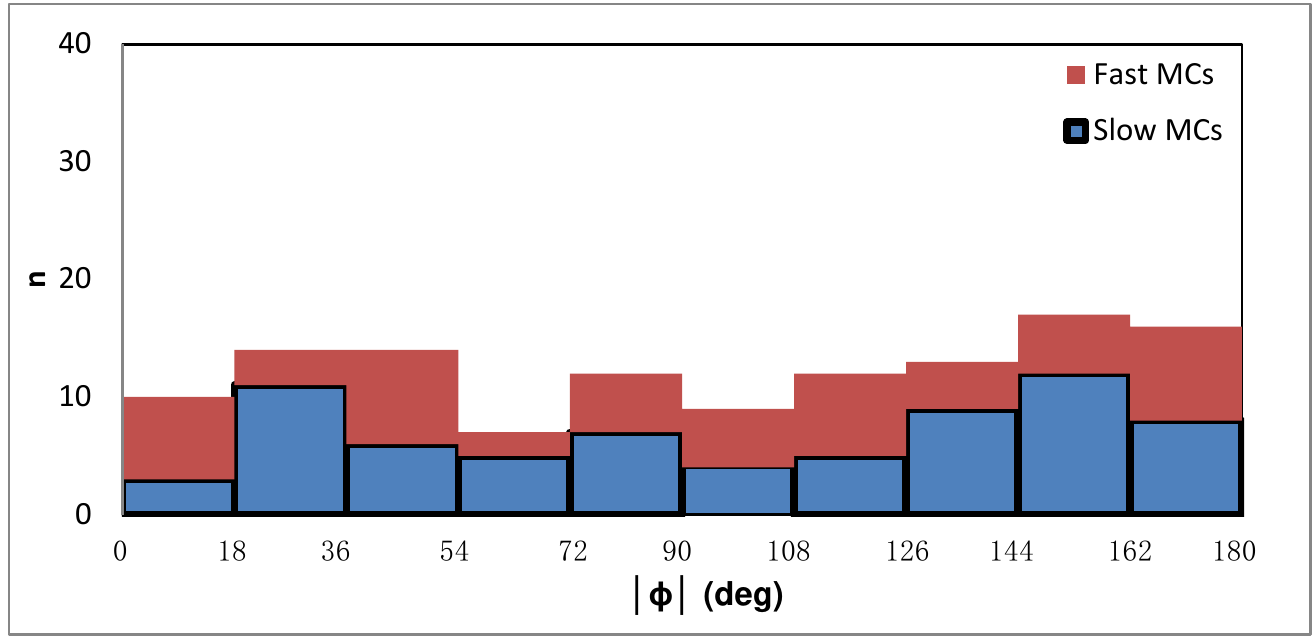}  \\
    \includegraphics[width=.45\textwidth]{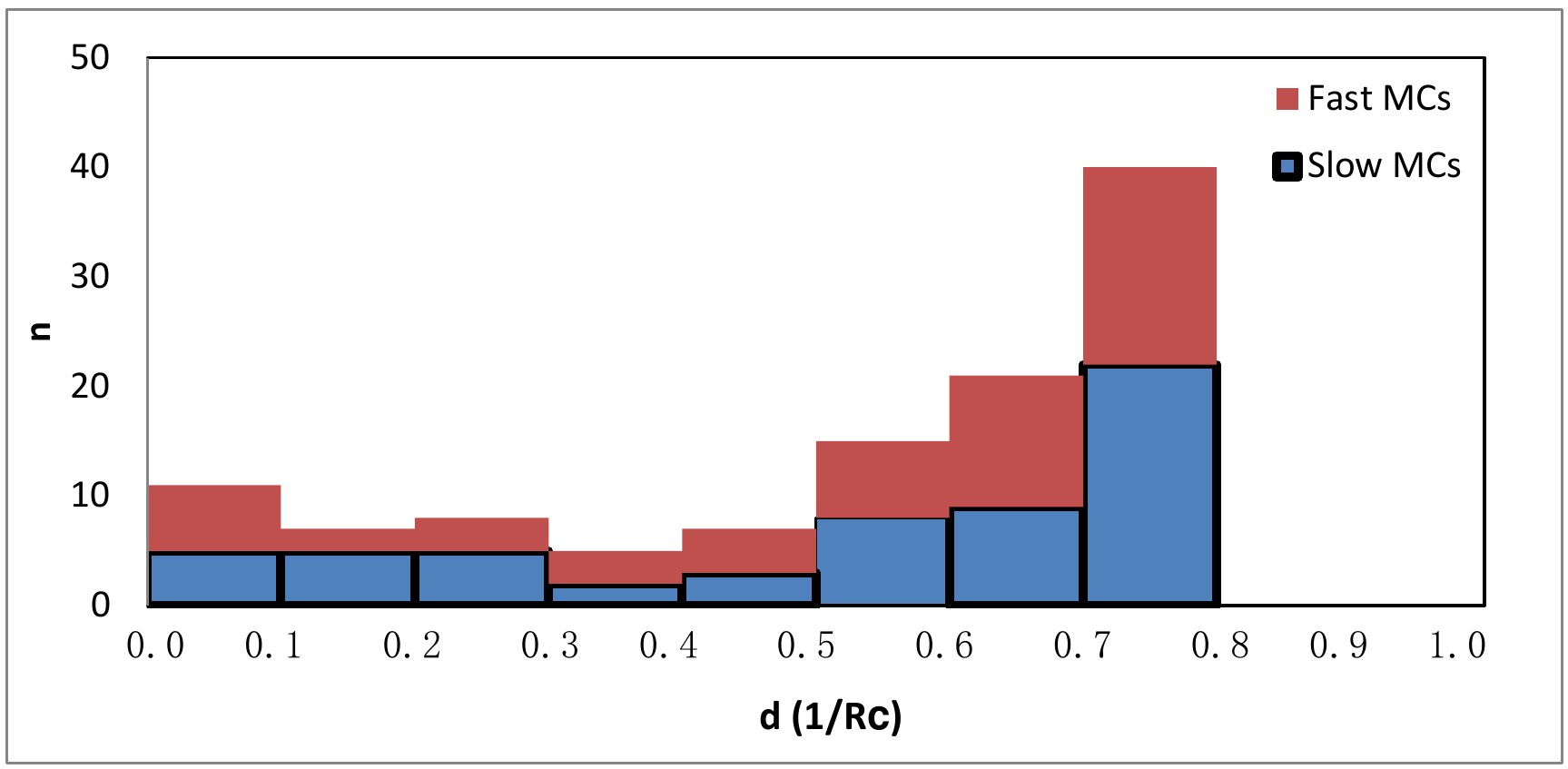} &
    \includegraphics[width=.45\textwidth]{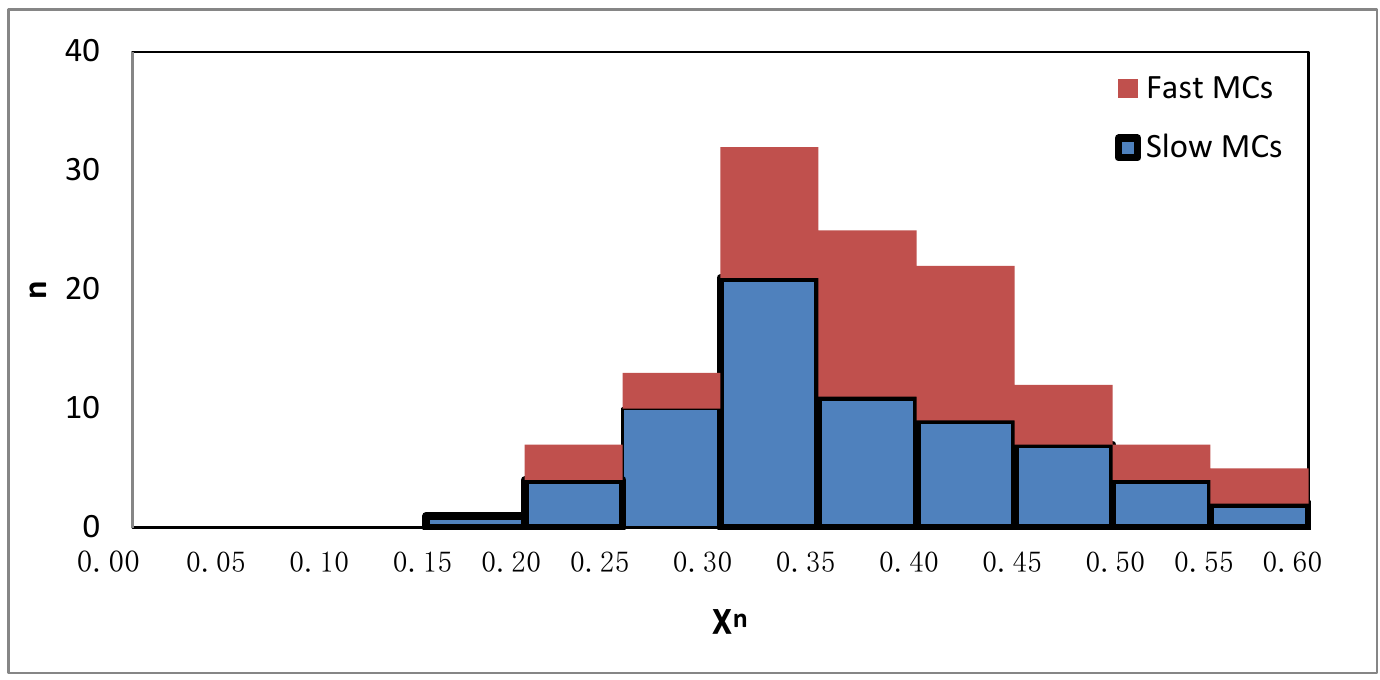} \\
  \end{tabular}
  \caption{Histogram distributions of MCs fit parameters. The red bars indicate fast MCs, the thick outlined bars in blue represent slow MCs.}
  \label{fig:mcfit}
\end{figure}

One of the solutions of Eq.(1) in the cylindrical geometry is Lundquist solution \citep{1950Ark.35...361}, in terms of axial [z], tangential [$\phi$], and radial [$\rho$] cylindrical components,
\begin{equation}\label{eq2}
  B_{\rho}=0,\qquad  B_{\phi}=HB_{0}J_{1}(\alpha\rho), \qquad  B_{Z}=B_{0}J_{0}(\alpha\rho)
\end{equation}
where $B_{0}$ is the field magnitude on the cylinder axis, $\rho$ is the distance from the axis, $H$ is the sign of the helicity, and $J_{0}$, and $J_{1}$ are zeroth and first-order Bessel functions, respectively. The constant value of $\alpha$ is derived from the radius of the cloud model cylinder ($R_{c}$), such that their product is the first zero of $J_{0}$. The fitting procedure described by \citet{1990JGR....9511957L} was applied to the analysis. The candidate events' possible boundaries were tested and chosen only when the boundaries were close to the best-fit boundaries and the fitting results were reasonably acceptable, i.e., normalized root-mean-square $\chi_{n}<0.6$ \citep{2018JGRA..123.3238W} and the closest distance of the spacecraft to the rope axis ($d$), namely impact parameter, in units of $R_{c}$, satisfies $d\leq0.8$. Finally, 64 ACE and 60 WIND observation events were selected. Table A1 lists the model fit parameters. Histograms of some fundamental parameters [$\bigtriangleup t$, $V_{rad}$, $R_{c}$, $B_{0}$, $\theta$, $\phi$, $d$,  $\chi_{n}$] are displayed in Figure \ref{fig:mcfit}. Definitions are given in the caption of Table A1. As shown by Figure \ref{fig:mcfit}, faster MCs have larger $B_{0}$ than slow ones. Moreover, our $\chi_{n}$ and $d$ tend to be large, but they are still within a reasonable range.

Comparing the events' boundaries with Lynch03' s, we found that 24/56 events have overlaps, but none of them are identical, which should cause the fitting parameters to have clear differences. We also compared our fitting results with KASI and \cite{2006AnGeo..24..215L}, whose boundaries are the same as ours. As listed in Table A2, take the events in 1998 as an example, most of the results can approximate theirs well.

\section{Method of inferring spatial position} \label{sec:method}

\begin{figure}[t]
\epsscale{0.65}
\plotone{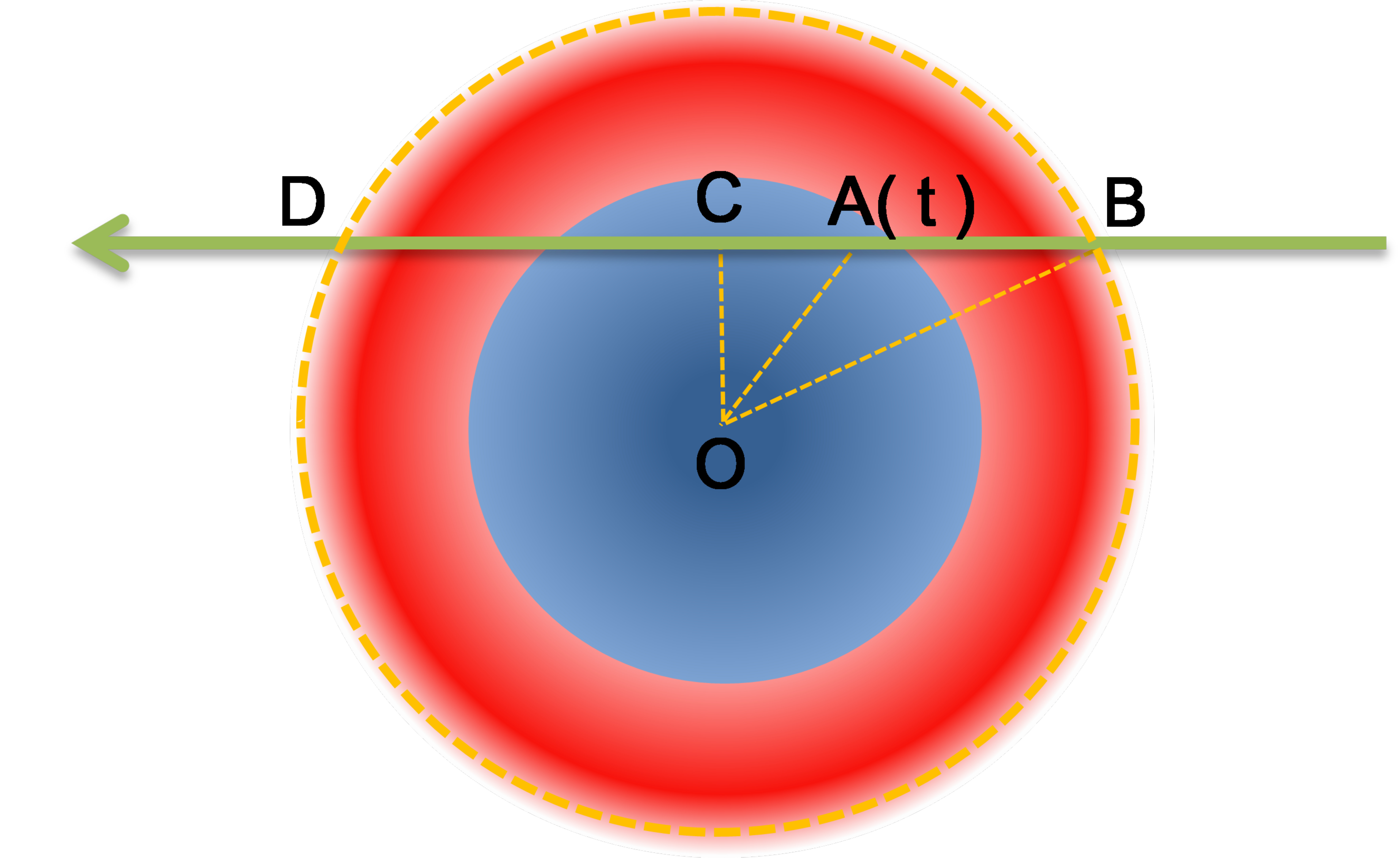}
\caption{Schematic drawing of the ACE spacecraft trajectory in the MC frame, showing the positions of a measured quantity during the MC event and the explanation of bimodal distributions of $\mathrm{\langle Q\rangle Fe}$ inside the MC. The orange dotted circle marks the boundary of the MC, the green arrow is ACE trajectory, A(t) is the measured quantity position, and red/blue denotes the high/low $\mathrm{\langle Q\rangle Fe}$ region.
\label{fig:cortoon}}
\end{figure}

We used Lynch03' s method (note that bins defined have some changes) to extract composition spatial structure from the magnetic field model. As shown in Figure \ref{fig:cortoon}, green arrows indicate the spacecraft trajectory and direction, and the circular region represents the MC cross-section. This figure describes the trajectory of the spacecraft in the cloud projected onto the plane perpendicular to the cloud axis. The ACE measured quantity has been coupled with the radial distance inside the model cylinder using the model geometry. We define $x$ as the normalized spatial position of the measured quantity.

\begin{equation}\label{eq:normalized}
  x = \frac{\mathrm{|OA|}}{\mathrm{|OB|}}
\end{equation}

In Eq.(3), $\mathrm{|OB|}$ is $R_{c}$, and $\mathrm{|OA|}$ can be easily calculated by $\mathrm{|AB|}$ (determined by ACE travelling time segment and flow velocity), $\mathrm{|OB|}$, and $\mathrm{|OC|}$ (that is $d$). Note that the distances $\mathrm{|OA|}$ are symmetric around the $\mathrm{|OC|}$ of the model fit. By applying this procedure, we can obtain the common MC structure among different sizes, and construct the statistical average of any measured quantity.

\section{Results} \label{sec:results}

\begin{figure}[b]
\centering
  \begin{tabular}{@{}cccc@{}}
    \includegraphics[width=.45\textwidth]{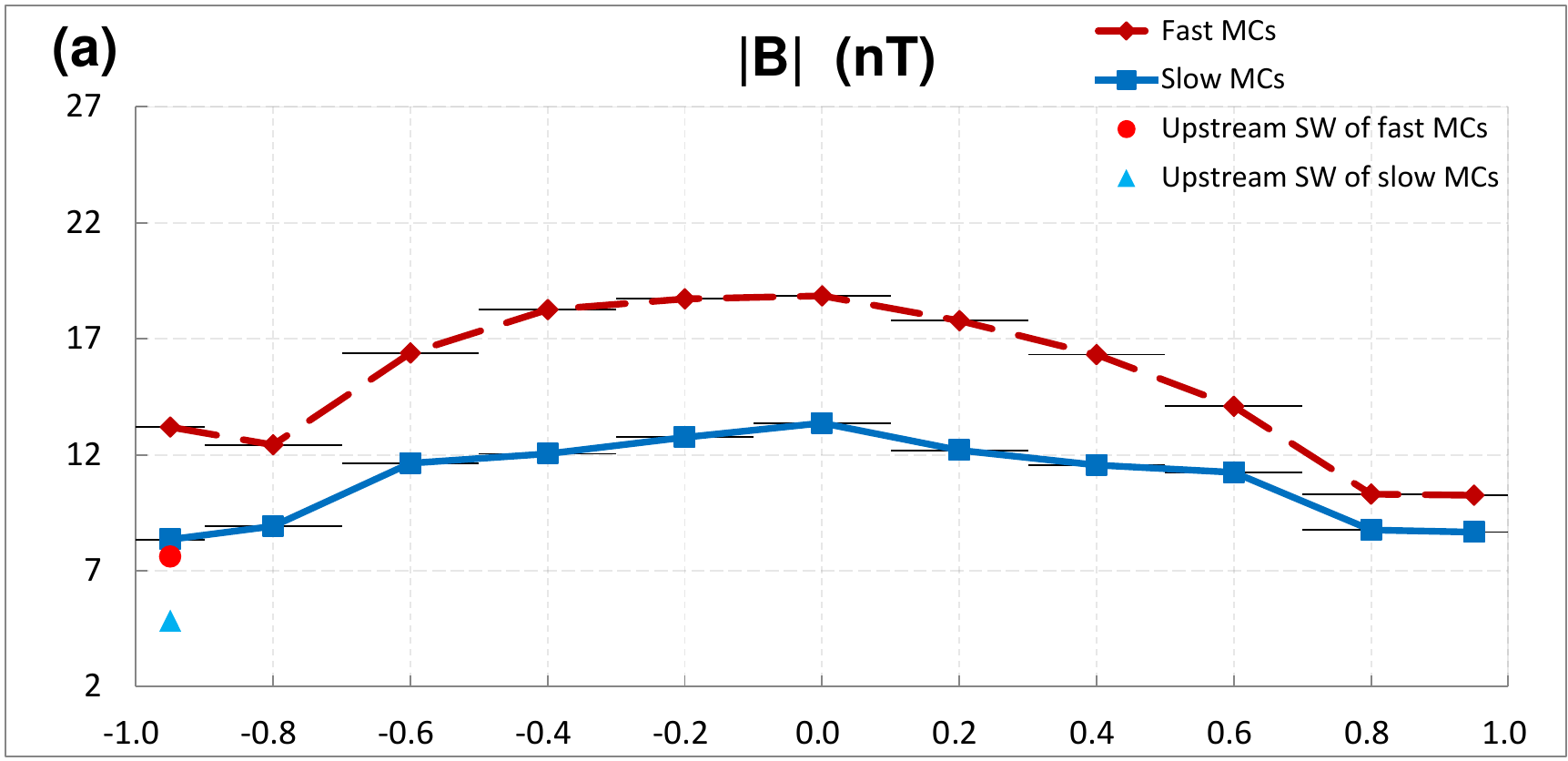} &
    \includegraphics[width=.45\textwidth]{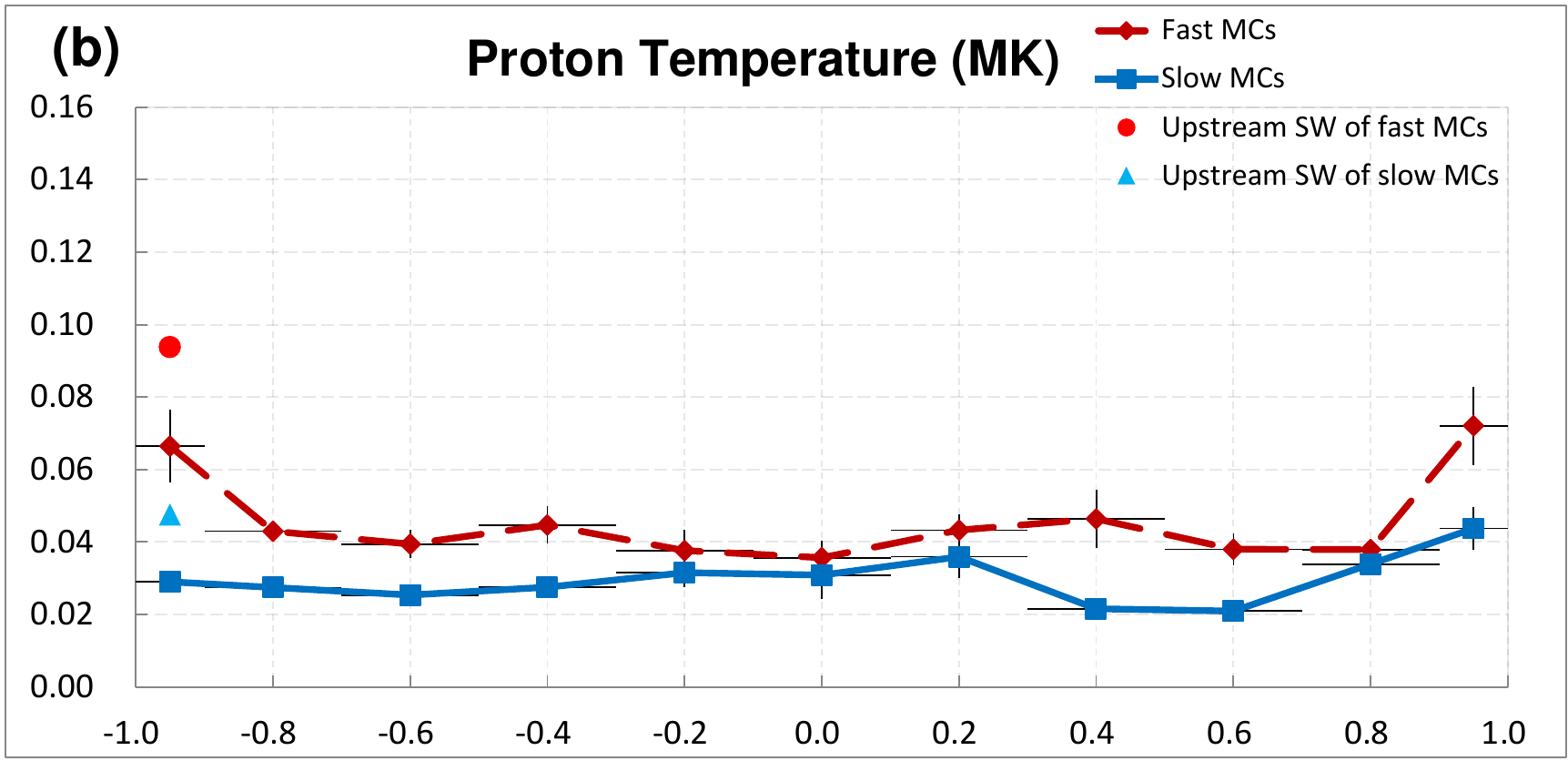}  \\
    \includegraphics[width=.45\textwidth]{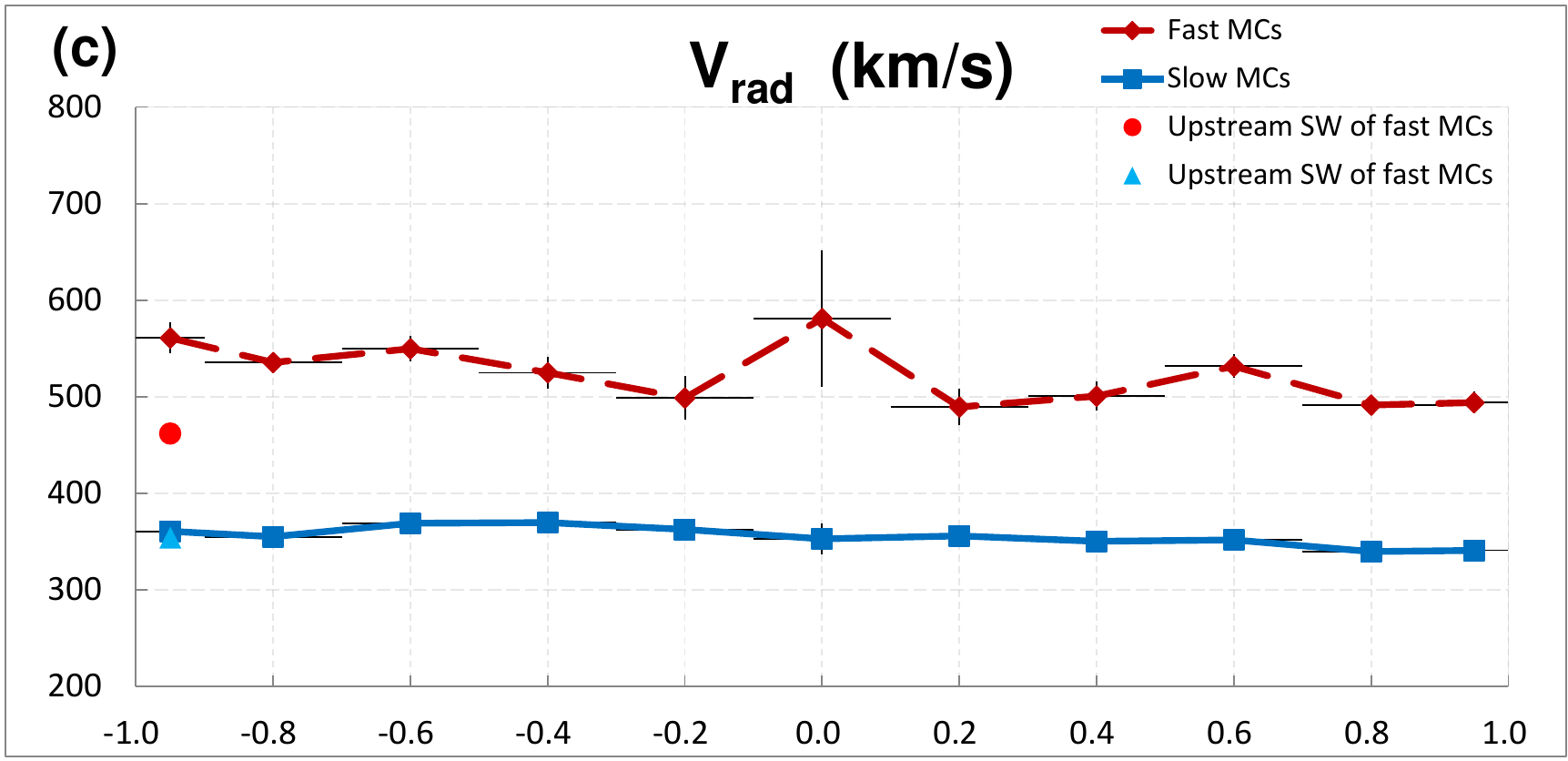} &
    \includegraphics[width=.45\textwidth]{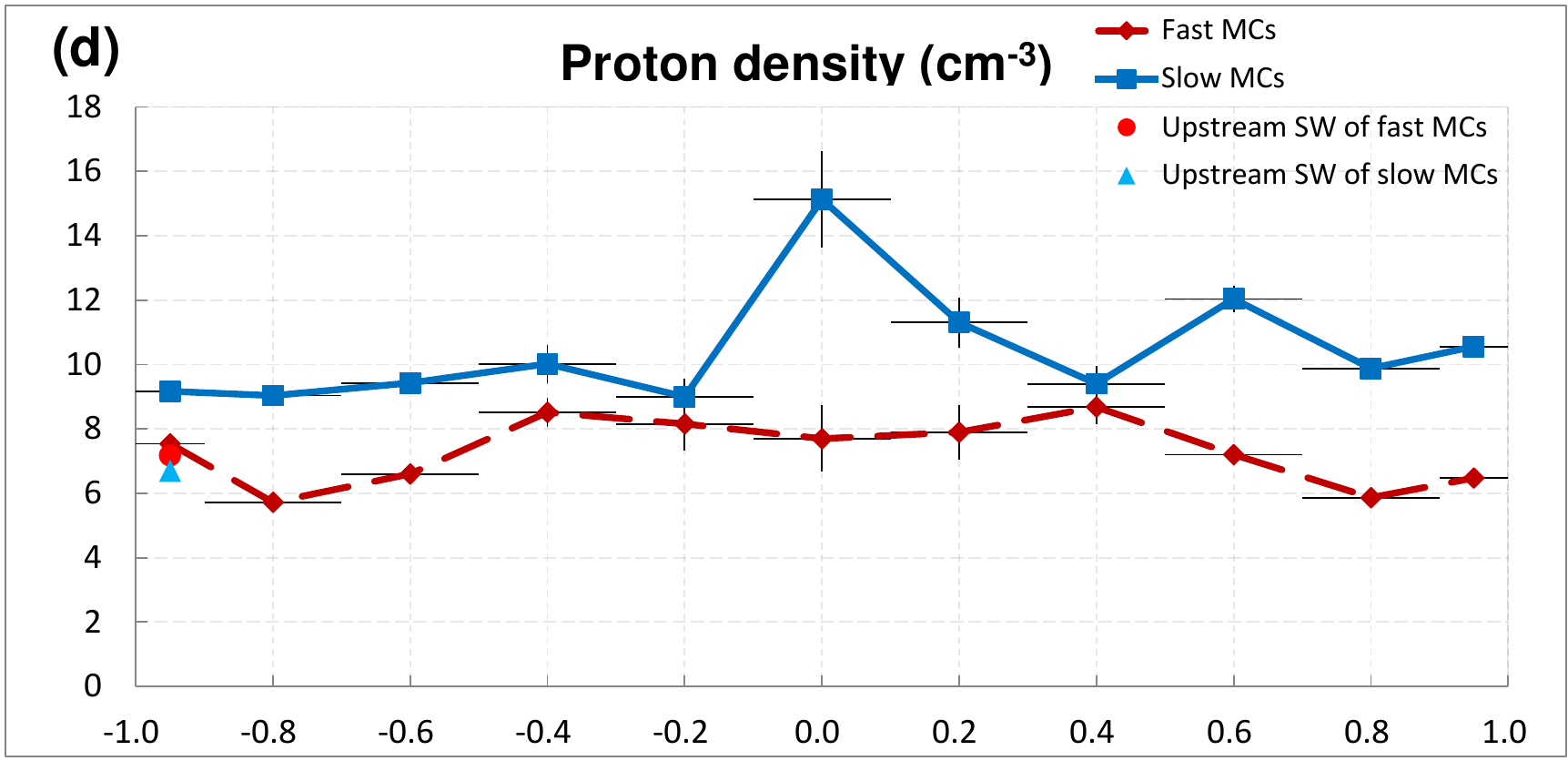}  \\
    \includegraphics[width=.45\textwidth]{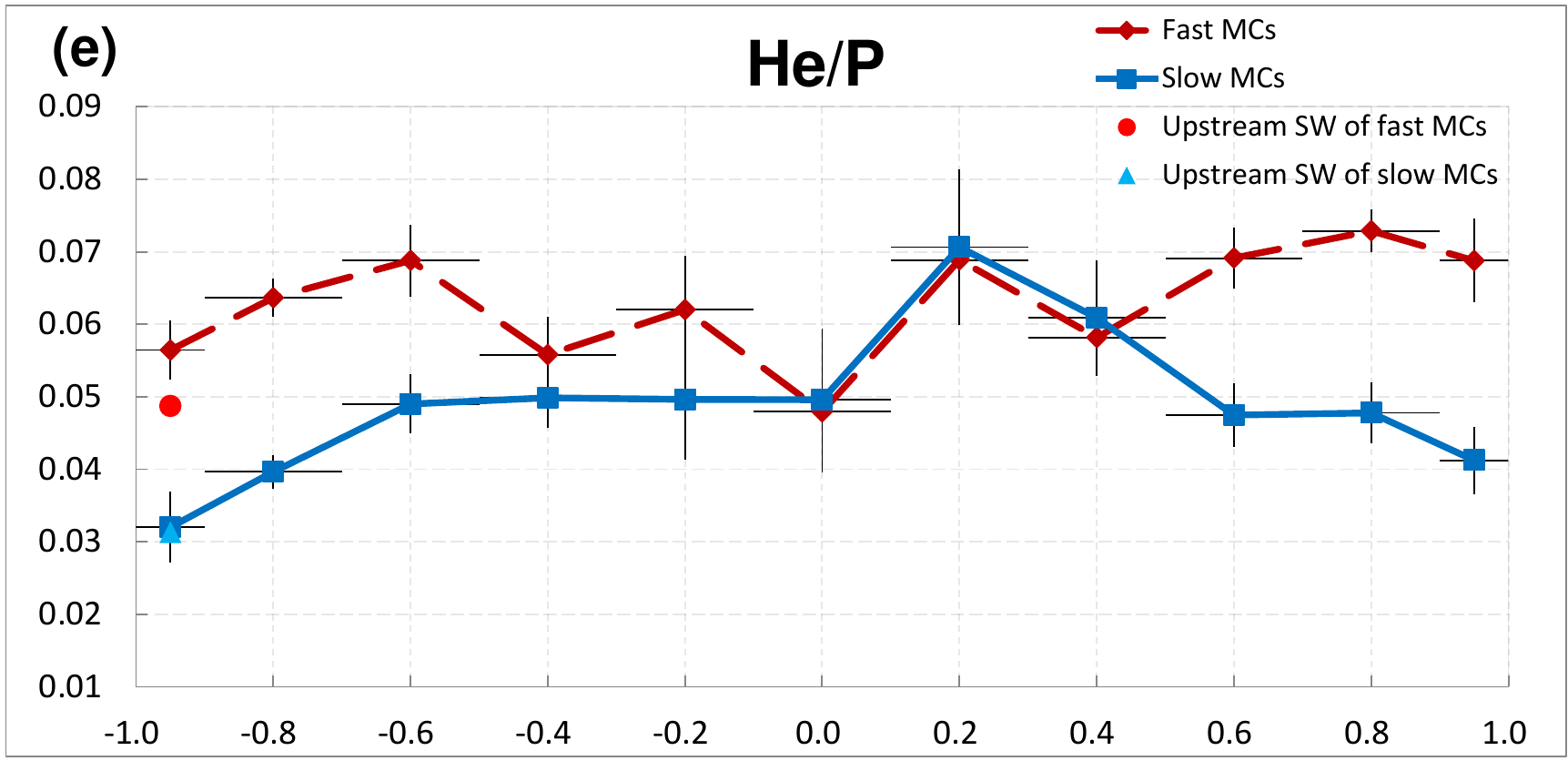} &
    \includegraphics[width=.45\textwidth]{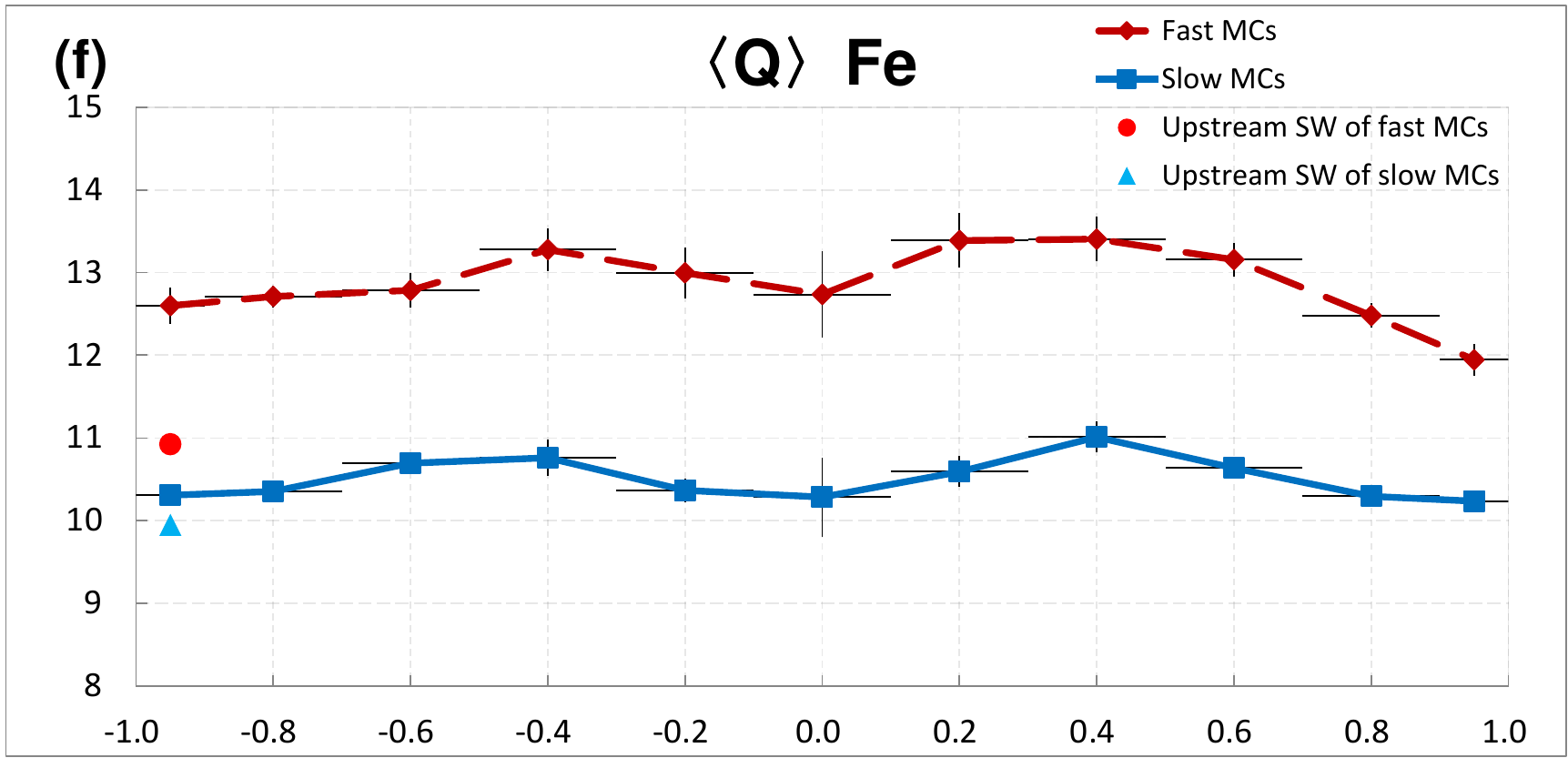} \\
    \includegraphics[width=.45\textwidth]{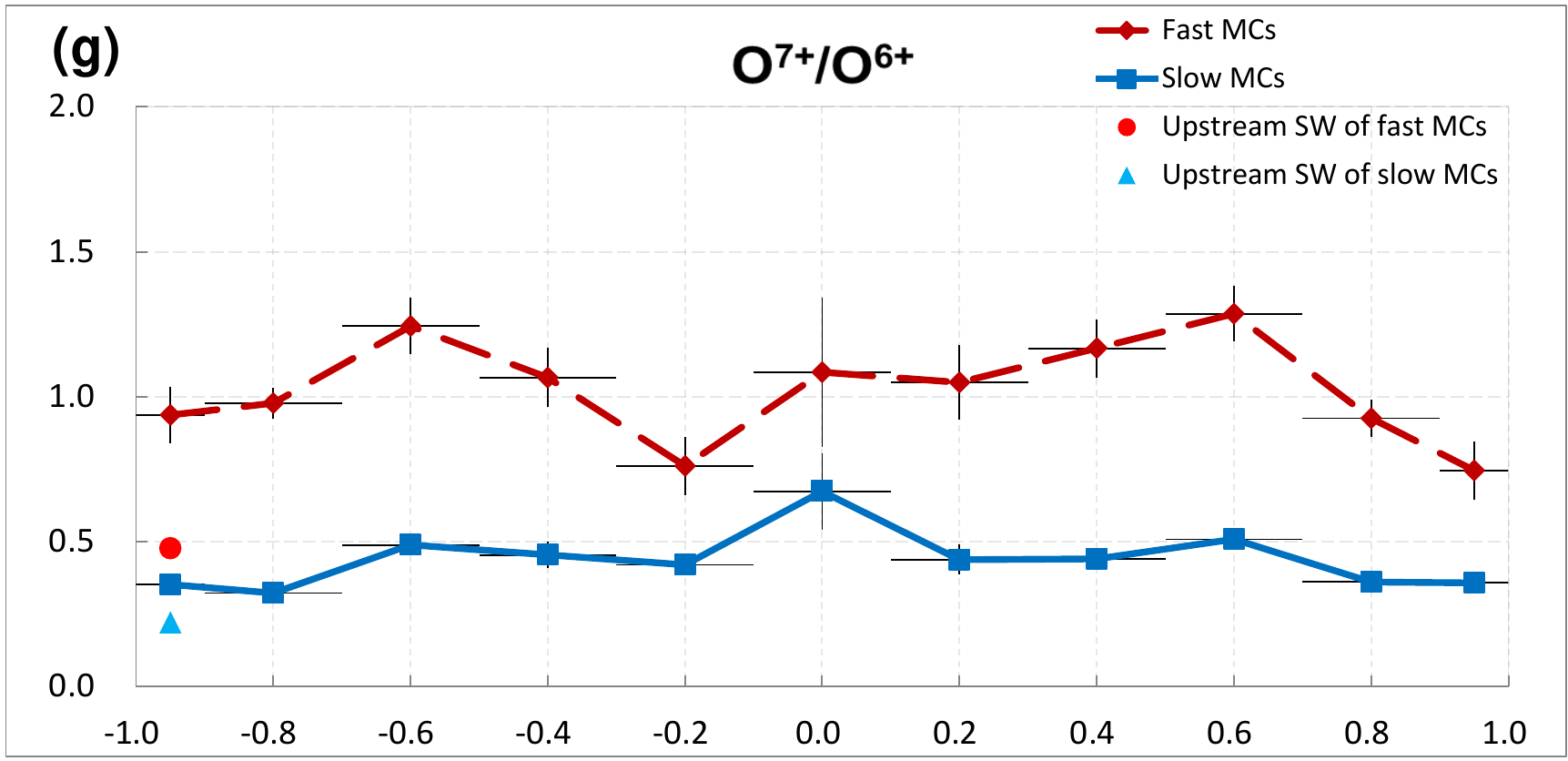} &
    \includegraphics[width=.45\textwidth]{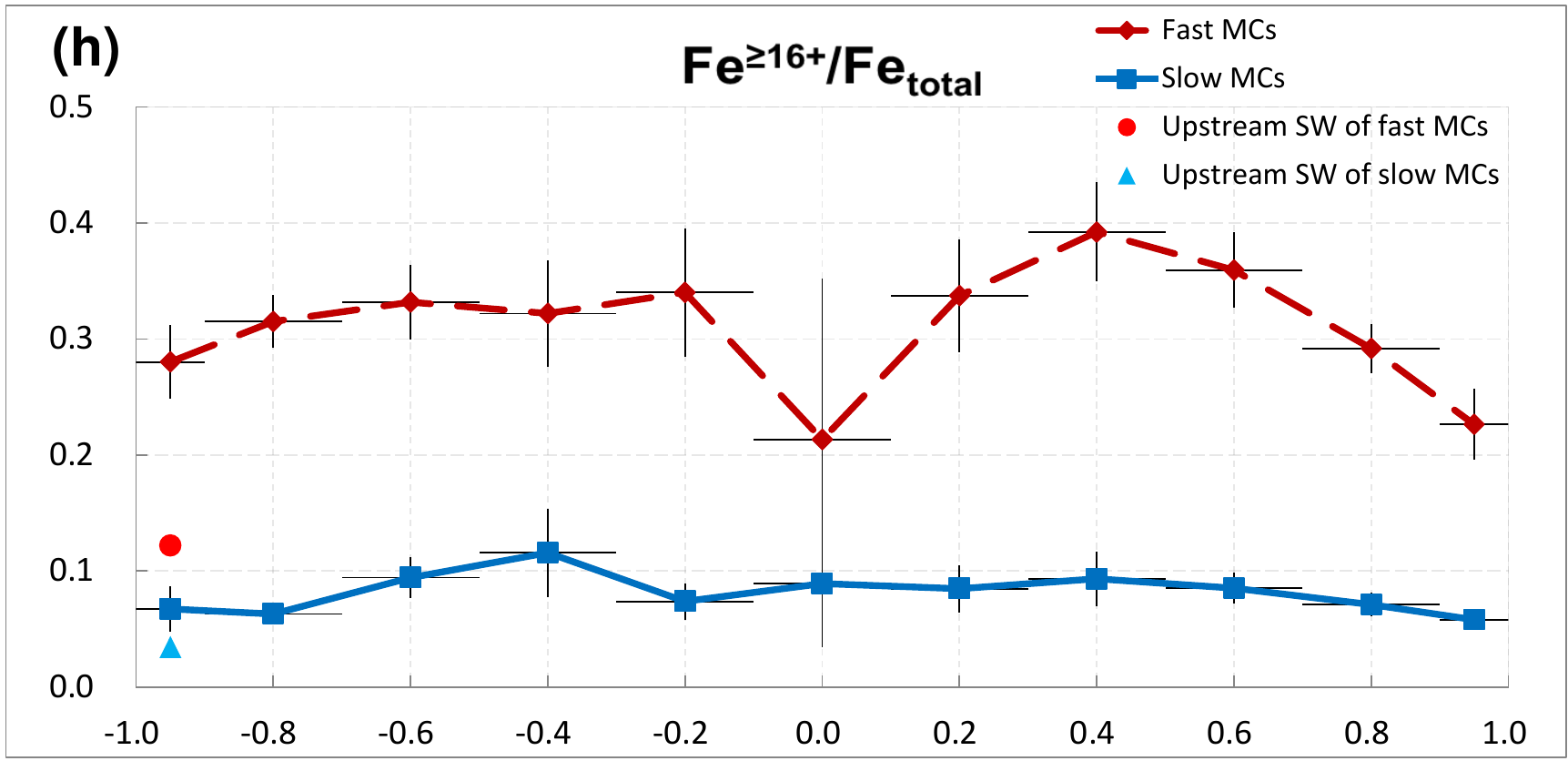} \\
    \includegraphics[width=.45\textwidth]{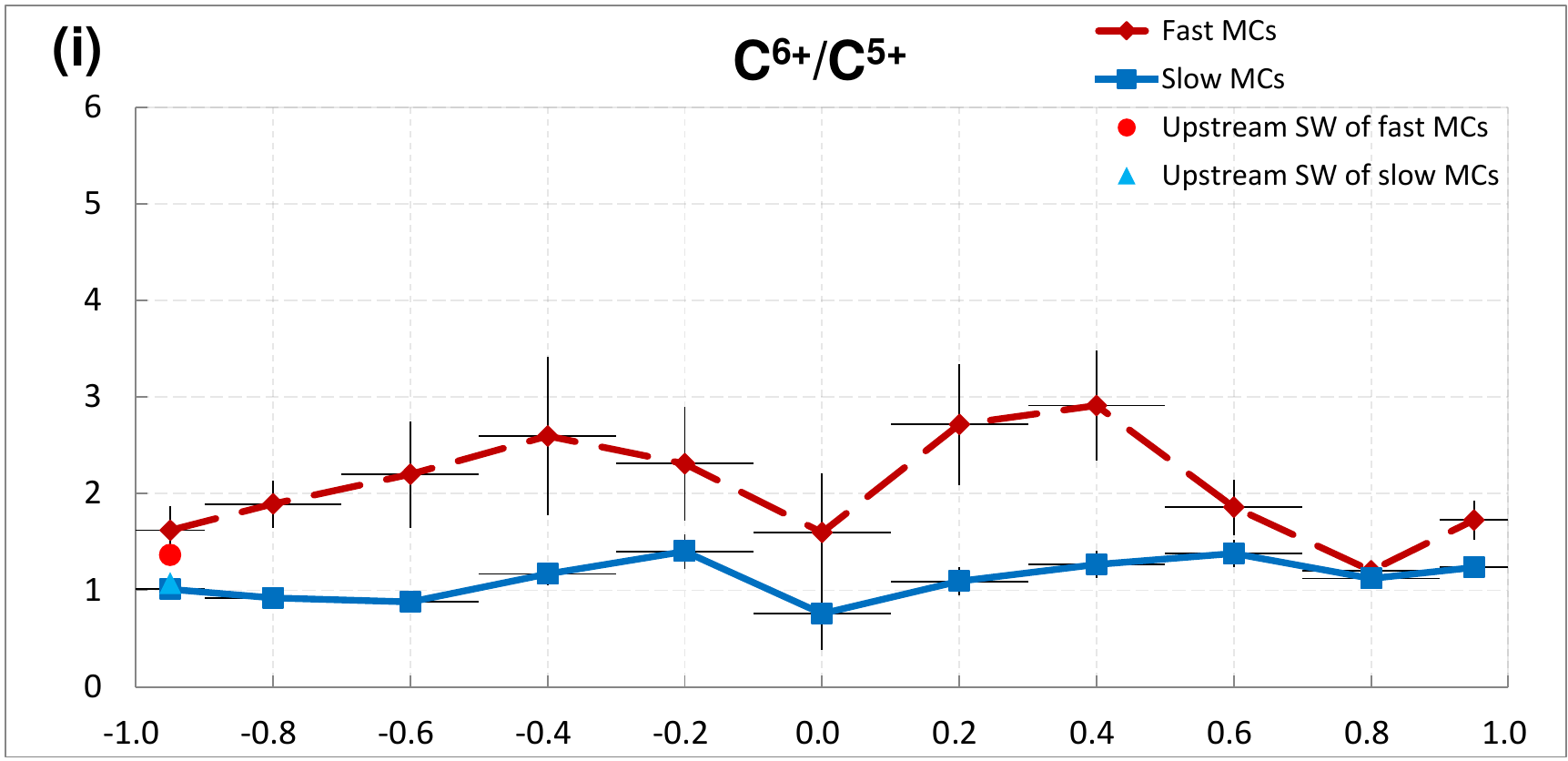} &
    \includegraphics[width=.45\textwidth]{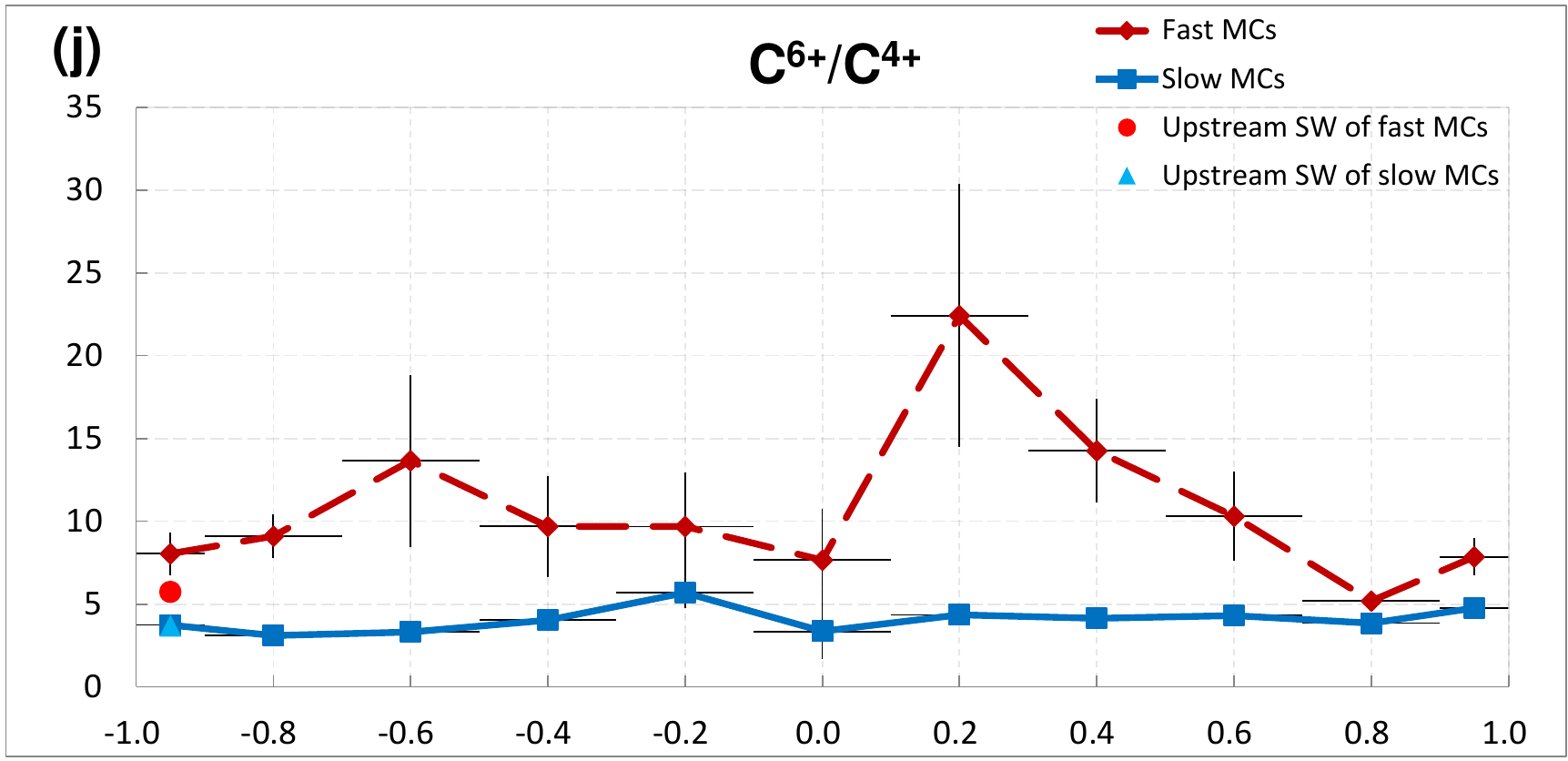}

       \end{tabular}
   \end{figure}

   \begin{figure}[t]
\centering
  \begin{tabular}{@{}cccc@{}}
    \includegraphics[width=.45\textwidth]{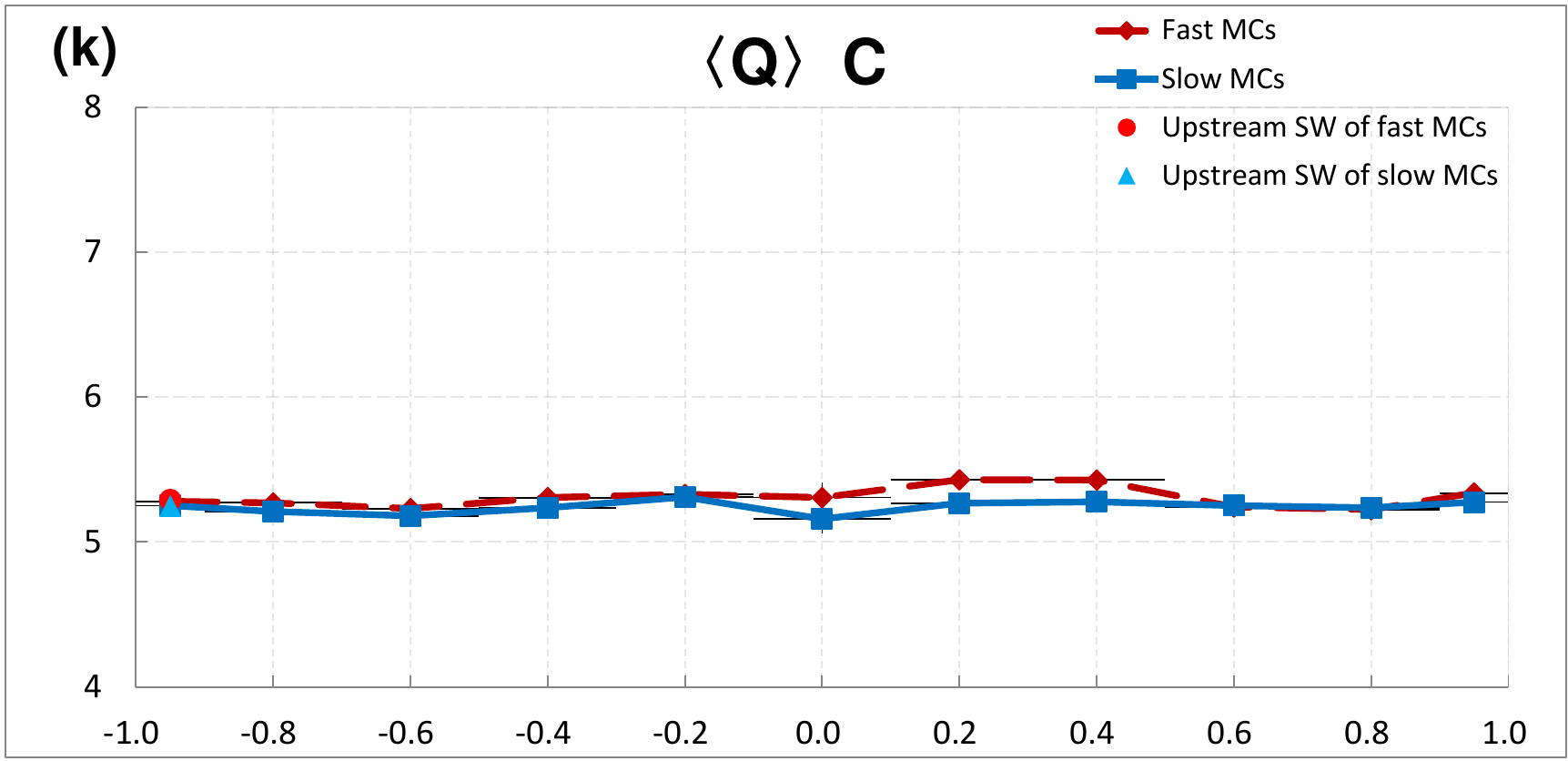} &
    \includegraphics[width=.45\textwidth]{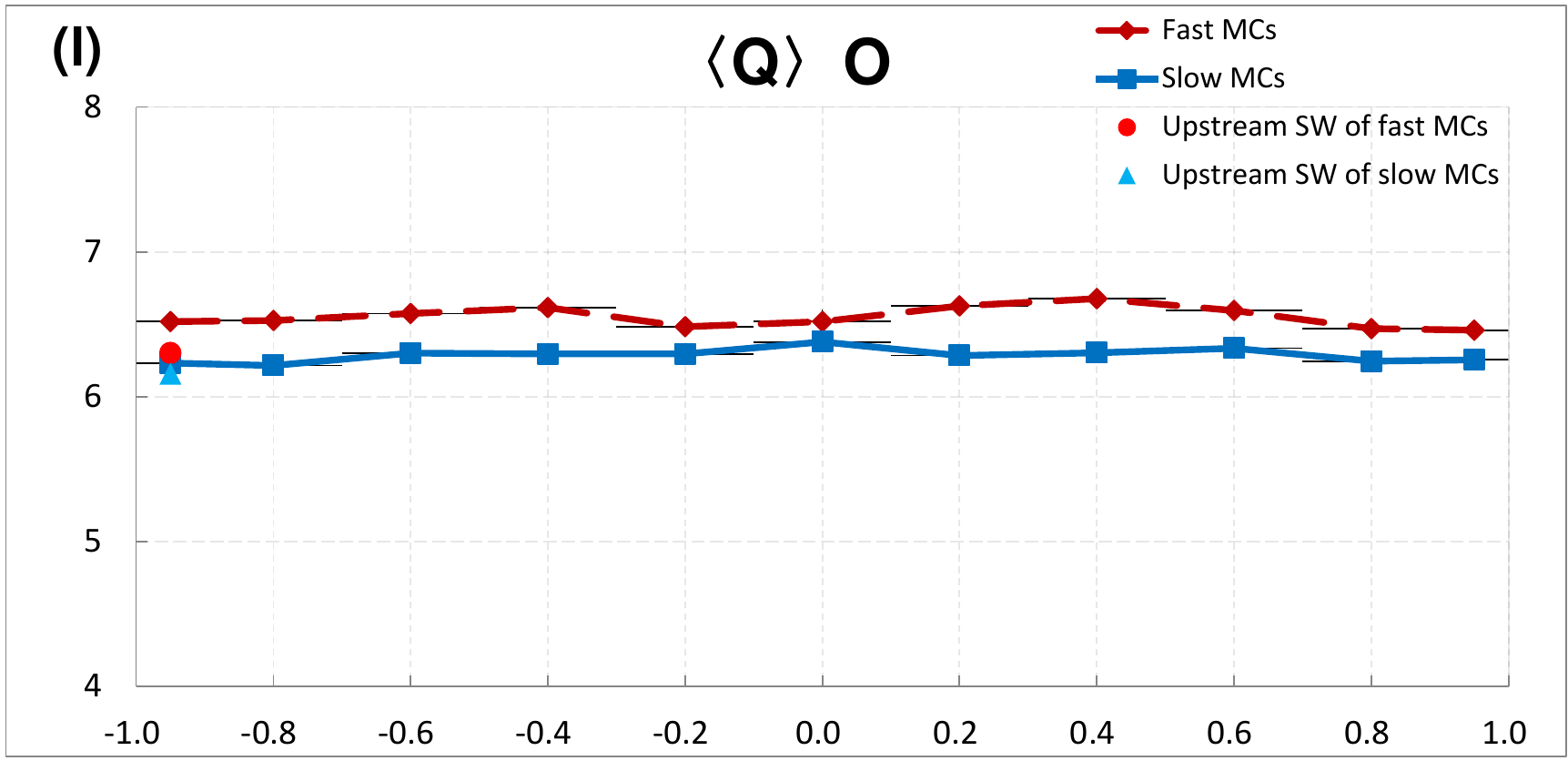} \\

    \includegraphics[width=.45\textwidth]{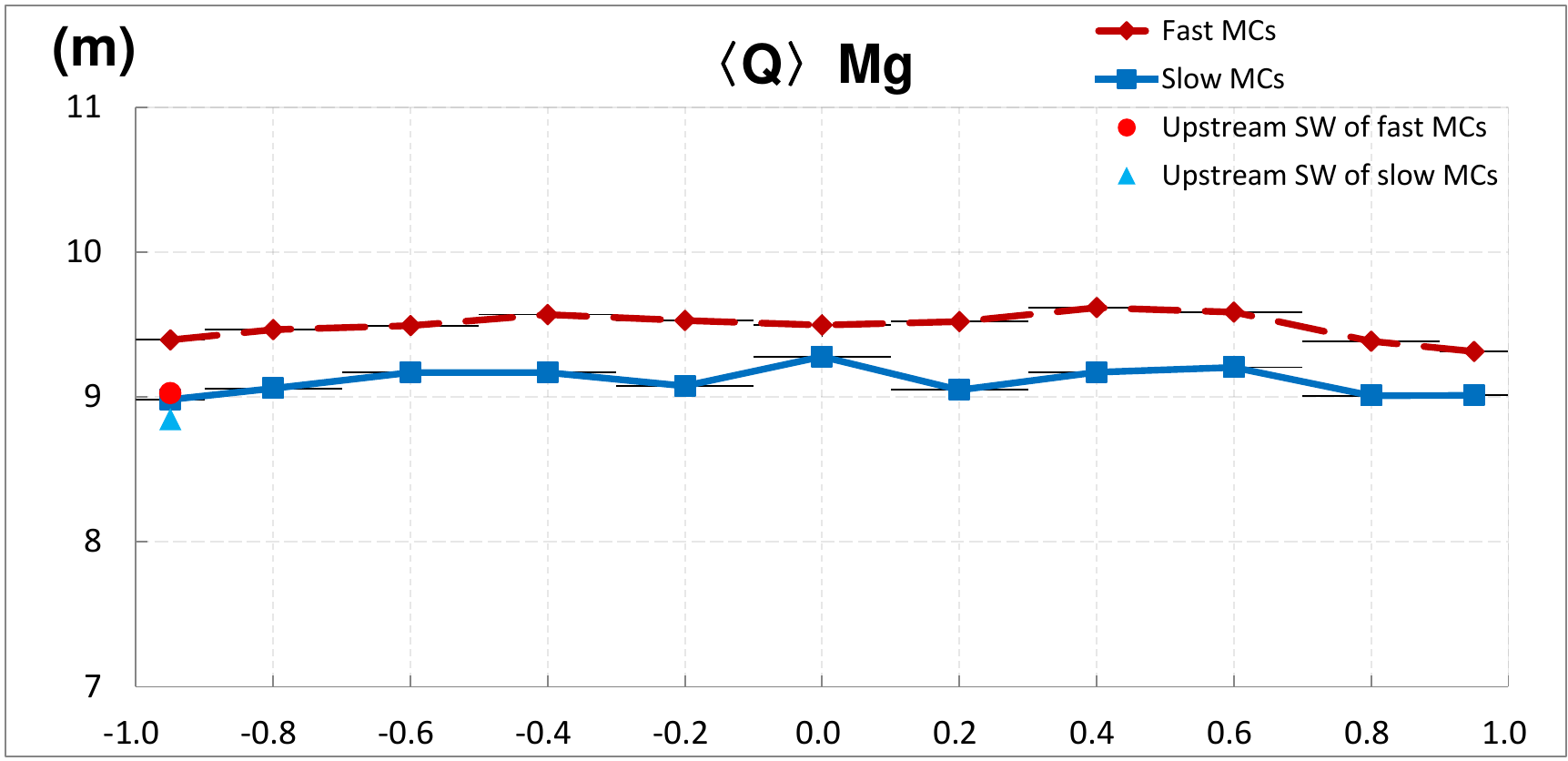} &
    \includegraphics[width=.45\textwidth]{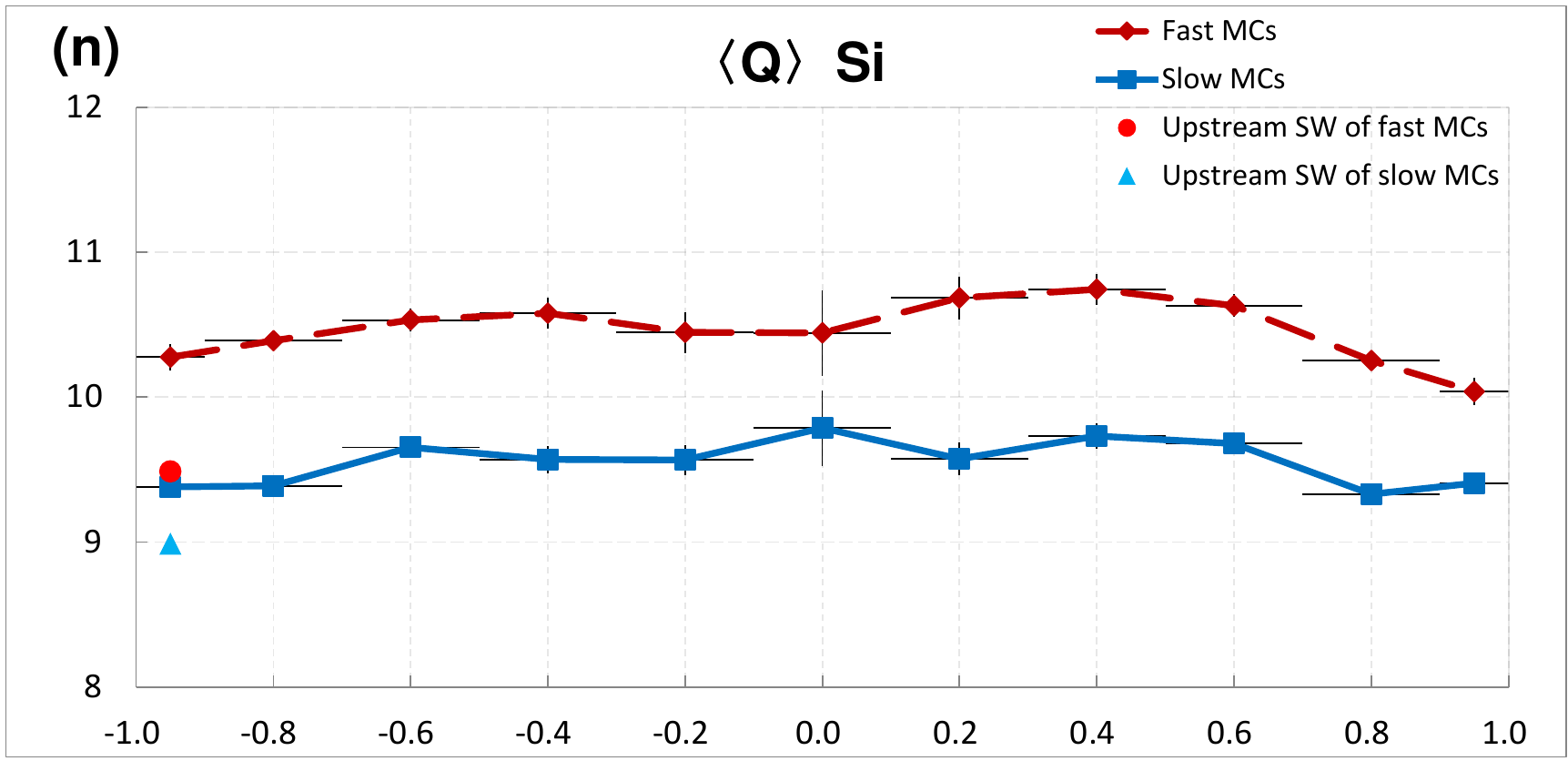}  \\
    \includegraphics[width=.45\textwidth]{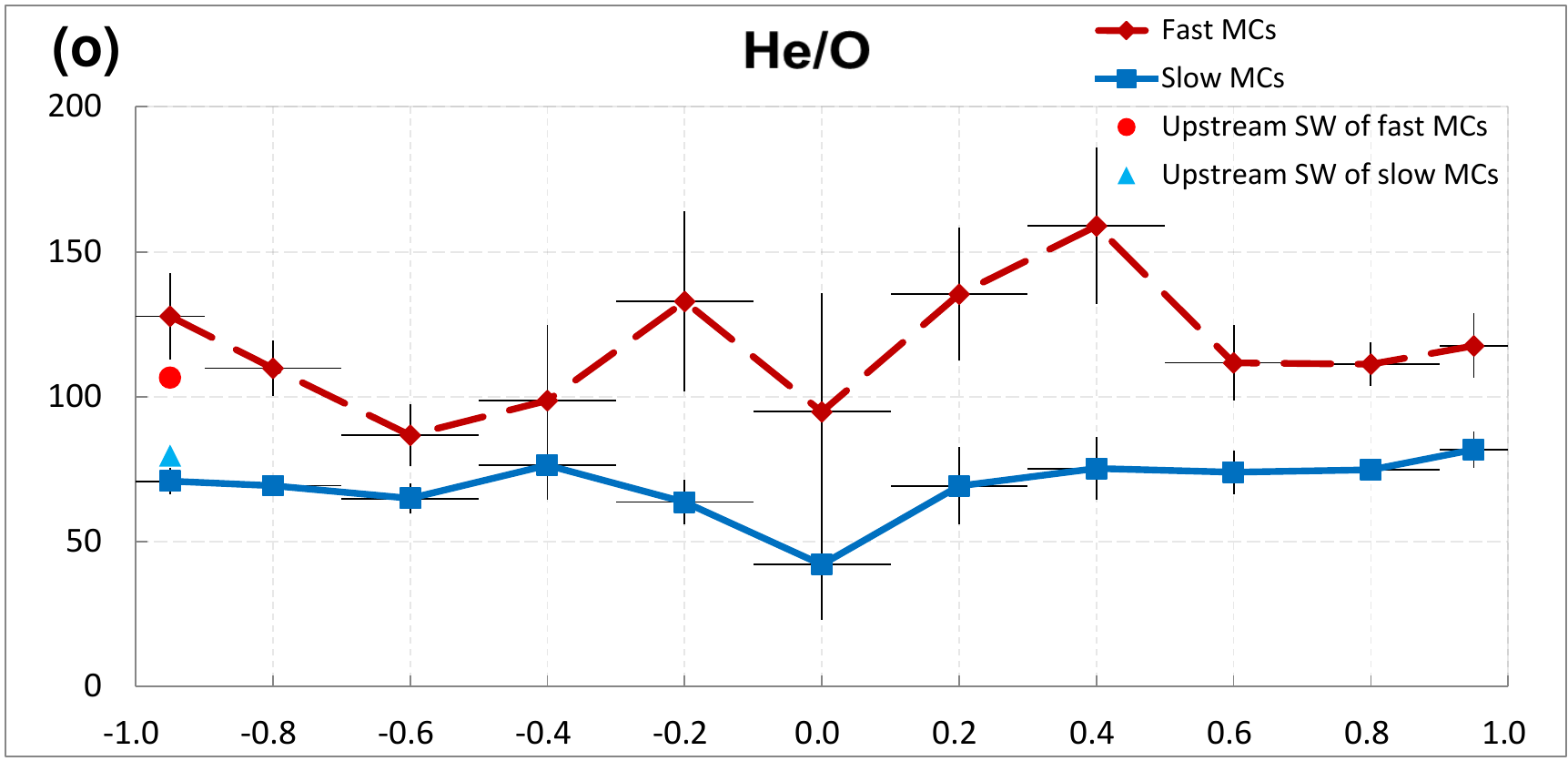} &
    \includegraphics[width=.45\textwidth]{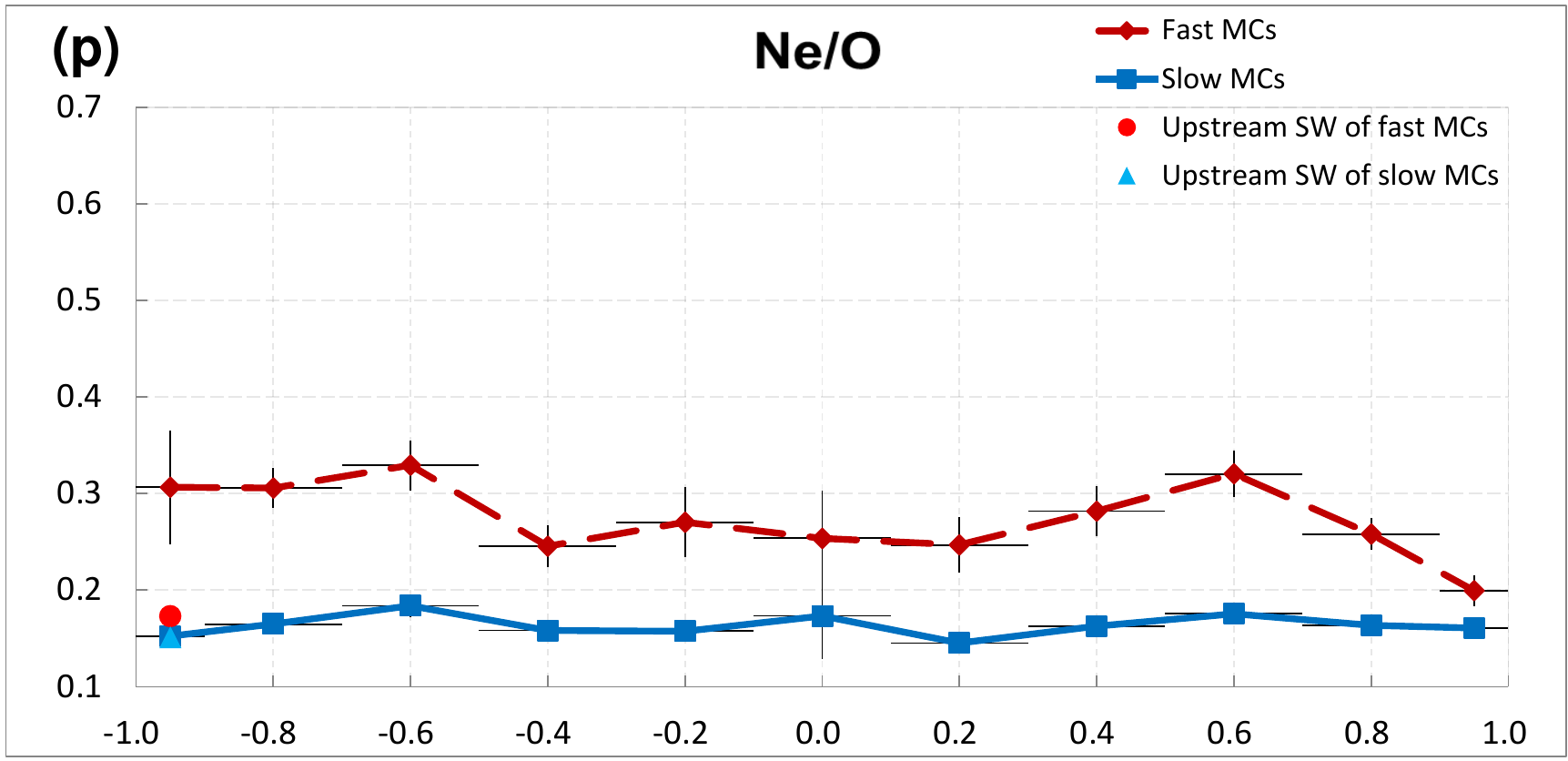}  \\
    \includegraphics[width=.45\textwidth]{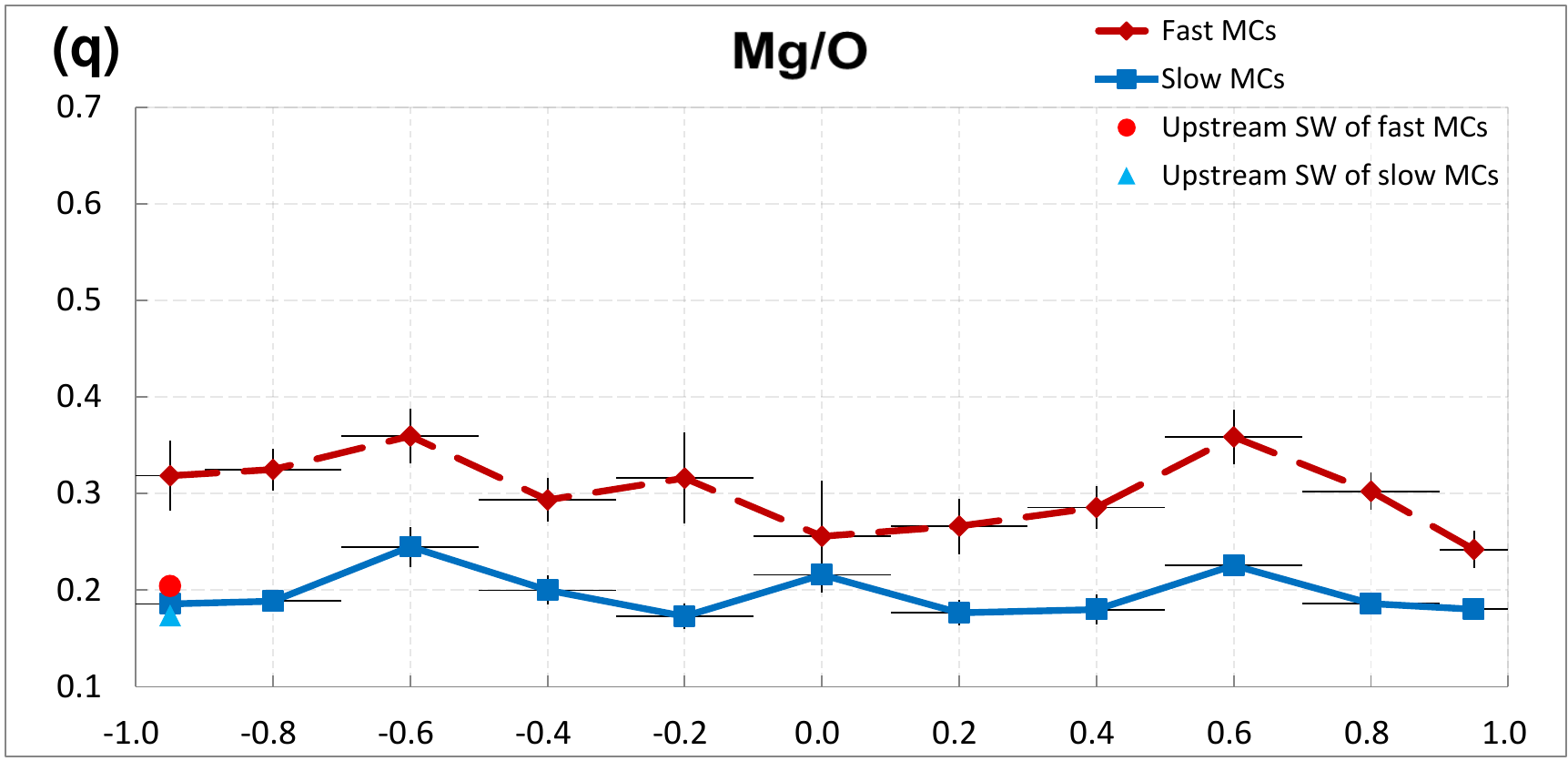} &
    \includegraphics[width=.45\textwidth]{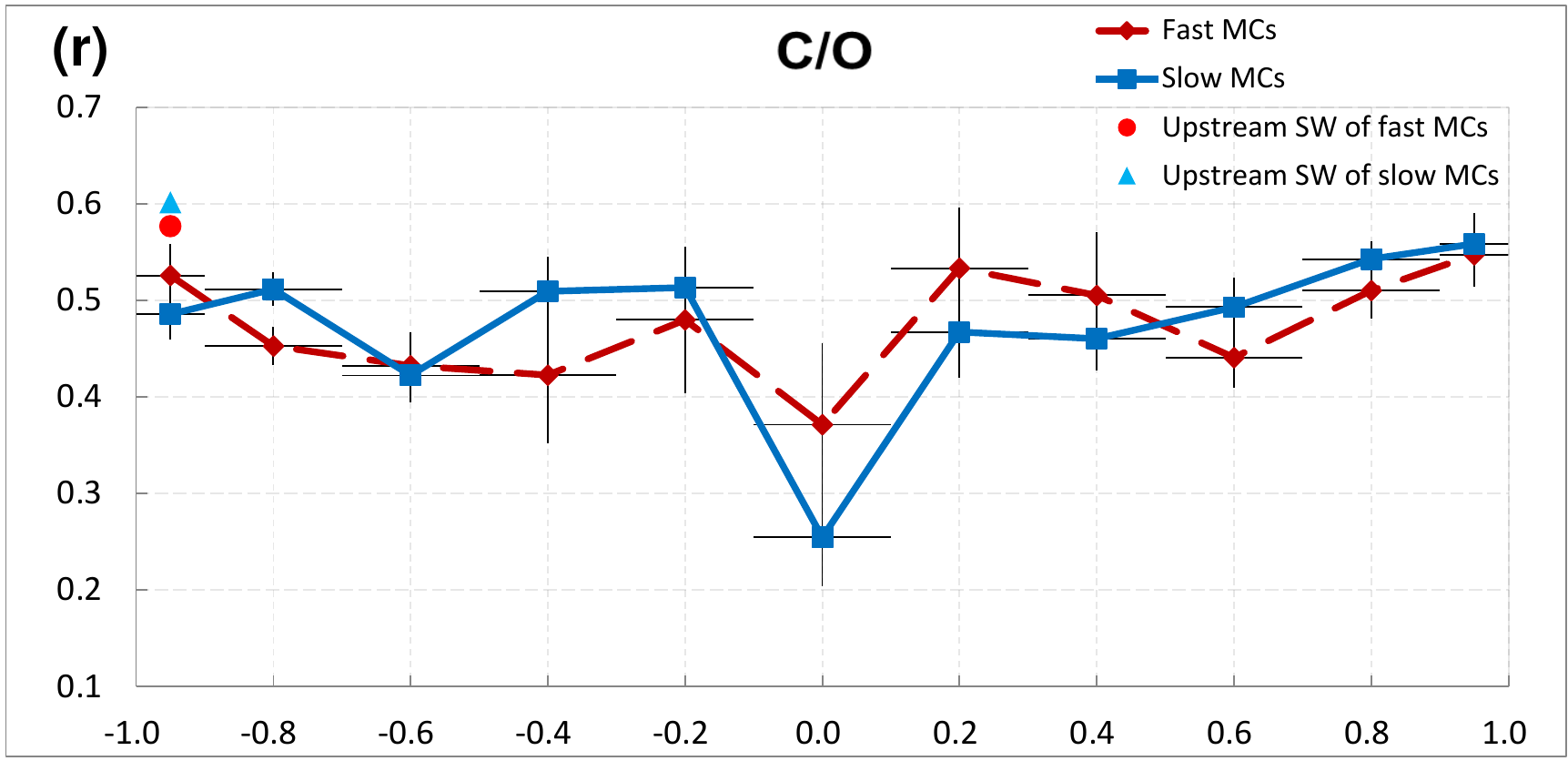} \\
    \includegraphics[width=.45\textwidth]{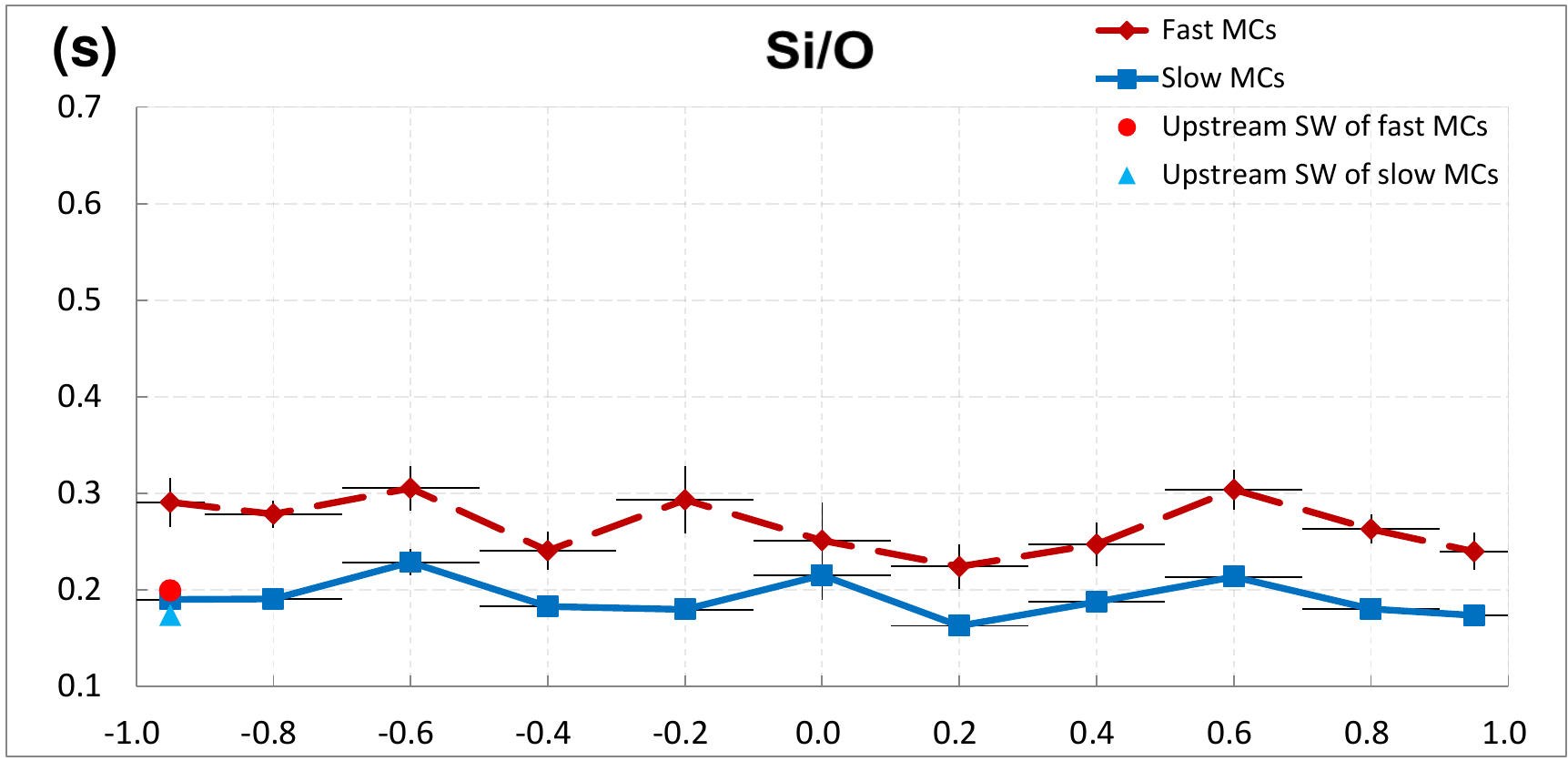} &
    \includegraphics[width=.45\textwidth]{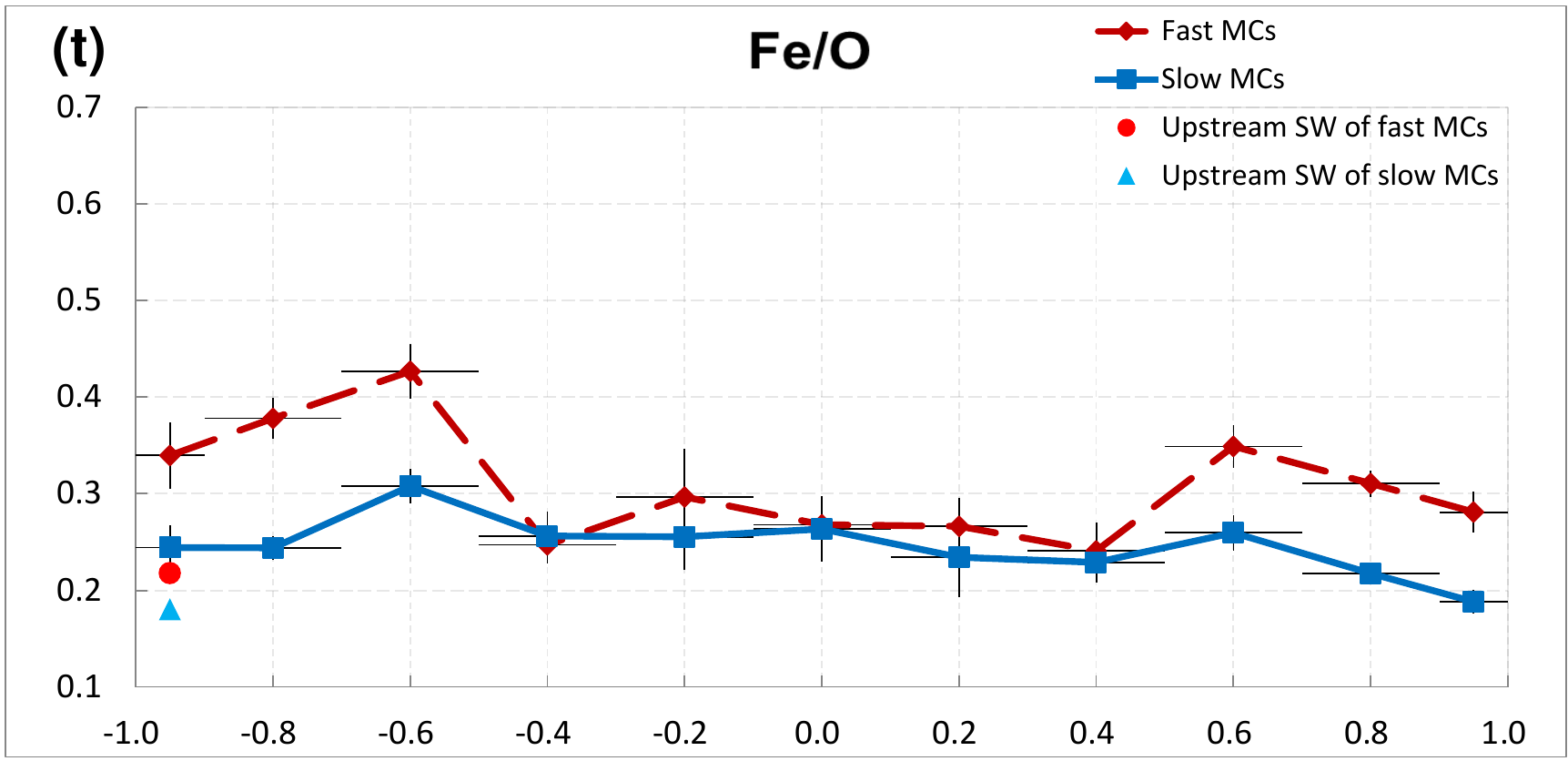} \\
  \end{tabular}
  \caption{Statistical distributions of plasma and composition inside MCs by ACE in 1998 - 2011. The red dashed lines are the quantity distributions inside fast MCs. The blue curves denote the quantity distributions inside slow MCs. Error bars represent the standard error of the mean in each bin. The red dots denote upstream SW mean value of fast MCs, the blue triangles denote the upstream SW mean value of slow MCs. Note that the edge bins span $-1.0$ to $-0.9$ and 0.9 to 1.0, respectively.
}
  \label{fig:plasma}
\end{figure}

In this paper, we made a comprehensive survey of the plasma and composition distributions inside 124 MCs during 1998 February - 2011 August. MCs have an average SW speed of 420 km/s near 1 AU \citep{1982JGR....87..613K}. Therefore, we classified the MCs into fast (55 MCs) and slow (69 MCs) types with a threshold of 420 km/s. After inferring the corresponding $x$ value of every measured quantity as section \ref{sec:method}, we divided the $x$ values into 11 bins, calculated the mean value in each bin, and let error bars denote standard error of the quantities within the bins. If the ACE spacecraft passes through MC along the green arrow in Figure \ref{fig:cortoon}, the quantity distributions inside MCs (Figure \ref{fig:plasma}) will be detected. The positive $x$-axis is the sunward side, whereas the negative $x$-axis is the earthward side. To compare with ambient SW status, SW mean value was calculated for all quantities. If there is a leading shock preceding MC, the calculated period is 12 hrs before the leading shock, otherwise, the period is 16 to 4 hrs before the leading edge of MCs. Four hours ahead of MCs here is to avoid the errors in MC boundary selections. All the times of the shocks and MC boundaries are shown with respect to ACE time. Results of all quantity distributions inside MCs are shown in Figure \ref{fig:plasma}, the related comments as follows:

(a) Magnetic field magnitude ($\left|B\right|$). Both fast and slow MCs show a domed-like profile. The fast MCs' profile is much higher and smoother than the slow MCs'. The slow MCs' profile is roughly symmetric about the mid-point, while the forward part of fast MCs is clearly higher than their rear part, which can be explained by the compression of upstream SW and violent radial expansion \citep{1993AdSpR..13...57O}. In addition, the leading edge bin (span $-1.0$ to $-0.9$) of the fast MCs shows a notable rise, the similar phenomenon is also found at proton temperature (Panel b) and proton density distributions (Panel d). The reason may be that there are some deviations in MC boundary determinations, thus the leading edges of fast MCs are mixed with contaminations from the ICME sheaths, which were typically generated by fast MCs.

(b) Proton temperature ($T_{p}$). Compared with the slow MCs, fast MCs show higher $T_{p}$, much hotter edges, which could result from the interaction with the ambient SW \citep{2007JGRA..112.8106G} or the other ICMEs \citep{2005ApJ...634..651L}.

(c) Radial velocity of SW ($V_{rad}$). As a result of severe and fluctuant expansion, it shows an uneven decrease throughout fast MCs, while the slight and smooth decline of slow MCs is consistent with their mild and steady expansion. The former is faster than the upstream SW, while the latter is about the same as the upstream SW at 1AU. Also, there is a protrusion in the center of fast MCs, the same signature can be seen in Figure 9 of Lynch03.

(d) Proton density ($N_{p}$). Contrary to $T_{p}$, fast MCs have lower $N_{p}$ than slow MCs. Both fast and slow MCs appear almost symmetric (coincide with the result of Lynch03) and denser center profiles, agree with the three-part CME structure. Besides, the $N_{p}$ of fast MCs is almost at the same level as upstream SW.

(e) He/P ratio. Both fast and slow MCs tend to increase from the leading to the trailing edge; this trend also appears in Figure 9 of Lynch03. In addition, many previous studies suggested that most of the ICMEs have high He/P. For instance, \citet{2004JGRA..109.9104R} used the criterion of He/P $> 0.06$ that are typically associated with ICMEs, \citet{2001JGR...10620957B} and \citet{2005JGRA..110.4103E} identified ICMEs with the criterion of He/P $>$ 0.08. As a subset of ICMEs, in our statistics, fast MCs are almost equal to 0.06, while slow MCs are lower than this criterion on average. It seems that He/P ratio tends to depleted if ICME has MC structure. In comparison to corresponding SW, both fast and slow MCs show a higher He/P ratio, which may be a consequence of reconnection-driven currents leading to chromospheric evaporation at the footpoints of the loops during CME eruption \citep{2003AIPC..679..604Z}.

(f) $\mathrm{\langle Q\rangle Fe}$. It is straightforward to analyze Fe charge states due to its large mass and high abundance in the solar wind. What's more, contrasting to C and O ions, which are easily affected by heating processes at the earlier stage, Fe ions freeze-in the altitude far from the Sun (e.g. $\mathrm{Fe^{8+}}$ $\sim$ $\mathrm{Fe^{12+}}$ freeze-in 1.3 $\sim$ 2.1 $R_{\odot}$ \citep{2003ApJ...582..467C}), thus they are sensitive to continuing heating in extended space. The most reliable indicator of Fe charge state behavior is $\mathrm{\langle Q\rangle Fe}$ \citep{2004JGRA..109.1112L}, providing an identifier of heating experience independent of expansion processes and heliocentric distance. Fast MCs have an apparently higher value than slow ones in $\mathrm{\langle Q\rangle Fe}$, as well as upstream SW. Both fast and slow MCs have a bimodal distribution, the peaks are symmetrically distributed at $x=\pm0.4$, and posterior peaks (13.40 for fast MCs, 11.01 for slow MCs) are slightly higher than those of the anterior ones (13.28 for fast MCs, 10.76 for slow MCs).

(g) $\mathrm{O^{7+}/O^{6+}}$ ratio. Since the dominant ions of oxygen, i.e. $\mathrm{O^{6+}}$ and $\mathrm{O^{7+}}$, freeze-in 1.0 $\sim$ 1.9 $R_{\odot}$ \citep{2003ApJ...582..467C}, $\mathrm{O^{7+}/O^{6+}}$ ratio is one of the diagnostics of low corona temperature. Panel (g) shows that fast MCs exhibit general bimodality in $\mathrm{O^{7+}/O^{6+}}$, but slow MCs distribution profile tends to be uniform.

(h) $\mathrm{Fe^{\geq16+}/Fe_{total}}$ ratio. The high iron charge states $\mathrm{Fe^{\geq16+}}$ (at least Fe \romannumeral17) representative of the materials heated by solar flares, with electron temperatures of up to 10 $\sim$ 20 MK \citep{2004JGRA..109.1112L}). Panel (h) shows that the enhancement of $\mathrm{Fe^{\geq16+}}$ within fast MCs is fairly pronounced, as well as the bimodality and further enhancement of posterior peak, whereas the distribution of slow MCs is nearly flat and there is no obvious posterior peak.

(j) \& (i) $\mathrm{C^{6+}/C^{4+}}$ and $\mathrm{C^{6+}/C^{5+}}$ ratio. All of them are indicators of the thermal environment in the source. The main carbon ionization states $\mathrm{C^{4+}}$, $\mathrm{C^{5+}}$, and $\mathrm{C^{6+}}$ freeze in 1.16 $\sim$ 1.26 $R_{\odot}$ \citep{2003ApJ...582..467C}, which is a relatively well-defined radial range than oxygen \citep{2012ApJ...744..100L}. They also exhibit bimodality for both fast and slow MCs, though $\mathrm{C^{6+}/C^{4+}}$ of slow MCs does not show obviously.

(k) - (n) Mean charge state distributions of carbon $\mathrm{\langle Q\rangle C}$, oxygen $\mathrm{\langle Q\rangle O} $, magnesium $\mathrm{\langle Q\rangle Mg} $, and silicon $\mathrm{\langle Q\rangle Si} $, respectively. Their profiles are not exactly the same. $\mathrm{\langle Q\rangle O}$ and $\mathrm{\langle Q\rangle Si}$ profiles show similar bimodal distributions to $\mathrm{\langle Q\rangle Fe}$, but the other two do not have such characteristics. This is because different elements correspond to different freeze-in altitudes, therefore, some of them are not affected by continuing heating along with a more extended space like Fe ions \citep[e.g.,][]{2010ApJ...722..625K,2011ApJ...730..103G,2012ApJ...760..105L}. In addition, combining Panel (f) and (k) - (n) and considering elements atomic number allowed an increase in mean charge state with the atomic number to be seen, which agree with characteristics in the solar wind and ICMEs \citep[e.g.,][]{1999JGR...10417005K,2012ApJ...761...48L,2012ApJ...760..141G}. Furthermore, if we sort the mean charge state of ions by the size of the difference between fast and slow MCs, that is $\mathrm{\langle Q\rangle Fe}$ $>$ $\mathrm{\langle Q\rangle Si}$ $>$ $\mathrm{\langle Q \rangle Mg}$ $>$ $\mathrm{\langle Q\rangle O}$ $>$ $\mathrm{\langle Q\rangle C}$. The order is also the same as the atomic number of elements from large to small. The cause is probably that the ions of the higher atomic numbers are easier to ionize to higher charge state by electron collisions or by direct heating.

(o) - (t) Elemental abundance of He, Ne, Mg, C, Si, and Fe relative to O respectively. Dominated by first ionization potential (FIP) fractionation or mass fractionation, they are calculated by summing over all charge states of each element. Compared with upstream SW, Mg/O, Si/O, and Fe/O in fast MCs exhibit abundance increase, all of them are low-FIP (FIP $\leq 10$ eV) elements relative to O. On the other hand, for high-FIP elements relative to O, Ne/O of fast MCs also shows significant enhancement, He/O shows less enhancement than upstream SW, while C/O shows apparent depletion. These characteristics of fast MCs are coincident with ICMEs explained by FIP fractionation \citep{2016ApJ...826...10Z}. Besides, slow MCs' elemental abundances are close to upstream SW, but much lower than fast MCs, which are either due to the intensive FIP effect that the pre-CME material of fast CME has experienced (different FIP effect implying different chromospheric structure or processes \citep{1995Sci...268.1033G}), or should be understood by mass fractionation \citep{2000JGR...10527239W}.

These results show that the magnetic field, plasma, and ionic composition distribution profiles inside fast and slow MCs appear in similarities and differences. In general, fast MCs tend to be enhanced and more fluctuant than slow ones.

\section{Discussion and Conclusions } \label{sec:conclusion}

In this study, we present the statistical plasma and composition distribution inside fast and slow MCs respectively, including He/P ratio, proton density, the magnetic field magnitude, proton temperature, average radial velocity, heavy-ion charge state of Fe, O, Si, Mg, and $\mathrm{O^{7+}/O^{6+}}$, $\mathrm{Fe^{\geq16+}/Fe_{total}}$, $\mathrm{C^{6+}/C^{4+}}$, and $\mathrm{C^{6+}/C^{5+}}$ ratios, as well as elemental abundance ratios of Ne/O, Mg/O, Si/O, C/O, and He/O. Our results indicate that the MCs have specific structures with statistical features in the plasma and composition quantities. The interior of fast MCs has enhanced ionic charge state distributions more than slow ones, which should be attributed to the fact that the origin of a heated coronal environment contains more energy at 1 AU. By comparing such enhancement of fast MCs for elemental species, an increase in the enhancement with atomic number is found. Additionally, for the distribution profiles, fast MCs are similar to slow MCs in the striking bimodality showed in $\mathrm{\langle Q\rangle Fe} $, $\mathrm{Fe^{\geq16+}/Fe_{total}}$, and $\mathrm{C^{6+}/C^{4+}}$ distributions. The difference is that the posterior peak of fast MCs is slightly higher than the anterior one, whereas two peaks are comparable in fast MCs. To the best of our knowledge, this is an unprecedented finding.

$\mathrm{\langle Q\rangle Fe}$ are good indications of coronal electron temperature \citep{2003JGRA..108.1239L}. A bimodal distribution of $\mathrm{\langle Q\rangle Fe} $ inside MCs indicates that the measured MCs contain a low ionized center and high ionized shell. \cite{2016ApJS..224...27S} consider that high temperature current sheet generates high charge state Fe ions, and fill in the corresponding layers of the flux rope, which may have been formed prior to the eruptions. In our statistics, bimodality is found in $\mathrm{\langle Q\rangle Fe} $ distributions inside both fast and slow MCs, implying that pre-existing flux rope is likely to be more common than we expected.

It is noteworthy that for the bimodal distribution of $\mathrm{\langle Q\rangle Fe}$ inside fast MCs, the peak close to the Sun is higher than the other one. $\mathrm{Fe^{\geq16+}/Fe_{total}}$, $\mathrm{O^{7+}/O^{6+}}$, and $\mathrm{C^{6+}/C^{4+}}$ also have similar characteristics. The possible cause is that the rear of fast flux rope CME faces the Sun. Based on the ``standard model'' for CME/flares \citep{2000JGR...105.2375L,2001JGR...10625199H}, when an eruption happens, the sunward side of flux rope connects to the current sheet, where electrons are accelerated and flow out into the rising flux rope subsequently. The electron beam is capable of ionizing surrounding  material by collisions \citep{1993ApJ...412..386M}. The sunward side flux rope ions get more high-energy electron collisions and would therefore elevate to higher charge states than the opposite side. Our results are also consistent with the previous observational study. \cite{2018ApJ...853L..18Y} analyzed an eruptive flare associated with a fast flux rope CME event using SDO/AIA data. As the temperature maps derived from the differential emission measure (DEM) showed (see Figure 4 in \cite{2018ApJ...853L..18Y}), there is a hot ring region inside flux rope during the eruption, and the bottom of the ring is hotter than the other regions. However, for slow MCs, the two peaks are comparable, possibly because of most slow MCs' counterparts are quiescent filament eruptions without flares.

\citet{2016JGRA..121.9316W} and \citet{2017ApJ...845..109Z} found that some MCs had a significant propagation velocity perpendicular to the radial direction, suggesting the direct evidence of the CME rotation in interplanetary space, and they inferred that a significant poloidal motion did exist in some MCs. Considering that the two peaks inside ionic charge state distribution of fast MCs are different in our statistics, it is likely that the propagation time of fast MCs in the interplanetary space is not enough to balance the ions in the front and rear before they are observed at 1 AU. Alternatively, the proportion of such poloidal motion in fast MCs is possibly too small to affect statistical results.

What determines the velocities of CMEs is still a question. In our explanation for the bimodal distribution of charge state inside MCs, magnetic reconnection is absolutely necessary. To be specific, the high-energy electron collisions of reconnection outflow cause a quite similar bimodality in $\mathrm{\langle Q\rangle Fe}$ distribution inside both fast and slow MCs. On the other hand, the ionic charge states of fast MCs are significantly higher than slow MCs, which means that the electron collisions and flare direct heating on the source region of fast CMEs are more pronounced with a higher reconnection participation level. This finding is in correspondence with the view of \citet{2005ApJ...634L.121Q} that CME velocities are proportional to the total reconnection flux estimated from the flare brightened region and extrapolation magnetic fields. It is reasonable that faster CME corresponds to the larger flare brightened region, which means that more fluxes are involved in the reconnection, causing higher energy electron collisions or higher temperature heating, then resulting in enhanced ionic charge state.

Finally, it is worth noting that our statistical results were based on the spatial structure which is derived from the application of the static, cylindrical, linear, force-free magnetic field model. Although this model was applied successfully to case studies and reproduced the general magnetic structure of the flux rope to some extent, it is based on some assumptions. Non-force-free fields, elliptical cross-section shape, and/or torus-shaped flux rope model may be closer to reality. Nevertheless, as commented on by Lynch03, complicated models bring diverse physical geometry of the flux rope, and it is difficult to construct an average profile with multiple events and perform statistic analysis.

\acknowledgments

This work is funded by the grants from the Strategic Priority Research Program of CAS with grant XDA-17040507, and the National Science Foundation of China (NSFC 11533009, 11973086). This work is also funded by the Project Supported by the Specialized Research Fund for Shandong Provincial Key Laboratory. In addition, we are also grateful to the One Belt and One Road Scientific Project of the West Light Foundation, CAS. The authors Y. Liu and Z. Z. Abidin would like to thank the University of Malaya Faculty of Science grant (GPF040B-2018) for support. All the ACE and WIND data are from NASA CDAWeb. We are grateful to the data provided by the NASA/GSFC. We are also grateful to the magnetic flux ropes list created by the Solar and Space Weather Research Group, Korea Astronomy and Space Science Institute. One of the authors, J. Huang, would like to thank Prof. Ilia Roussev for his stimulating discussion. Anonymous reviewer and editors' comments that were valuable for improving our manuscript are acknowledged.


\begin{thebibliography}{}

\bibitem[Bemporad et al.(2006)]{2006ApJ...638.1110B} Bemporad, A., Poletto, G., Suess, S.~T., et al.\ 2006, \apj, 638, 1110
\bibitem[Bothmer \& Schwenn(1998)]{1998AnGeo..16....1B} Bothmer, V., \& Schwenn, R.\ 1998, Annales Geophysicae, 16, 1
\bibitem[Burlaga et al.(1981)]{1981JGR....86.6673B} Burlaga, L., Sittler, E., Mariani, F., et al.\ 1981, \jgr, 86, 6673
\bibitem[Burlaga et al.(1982)]{1982GeoRL...9.1317B} Burlaga, L.~F., Klein, L., Sheeley, N.~R., et al.\ 1982, \grl, 9, 1317
\bibitem[Burlaga(1988)]{1988JGR....93.7217B} Burlaga, L.~F.\ 1988, \jgr, 93, 7217
\bibitem[Burlaga et al.(2001)]{2001JGR...10620957B} Burlaga, L.~F., Skoug, R.~M., Smith, C.~W., et al.\ 2001, \jgr, 106, 20957
\bibitem[Chen et al.(2006)]{2006A&A...456.1153C} Chen, A.~Q., Chen, P.~F., \& Fang, C.\ 2006, \aap, 456, 1153
\bibitem[Chen(1996)]{1996JGR...10127499C} Chen, J.\ 1996, \jgr, 101, 27499
\bibitem[Cheng et al.(2013)]{2013ApJ...763...43C} Cheng, X., Zhang, J., Ding, M.~D., et al.\ 2013, \apj, 763, 43
\bibitem[Cheng et al.(2014)]{2014ApJ...789...93C} Cheng, X., Ding, M.~D., Zhang, J., et al.\ 2014, \apj, 789, 93
\bibitem[Cheng et al.(2014)]{2014ApJ...789L..35C} Cheng, X., Ding, M.~D., Zhang, J., et al.\ 2014, \apjl, 789, L35
\bibitem[Chen et al.(2003)]{2003ApJ...582..467C} Chen, Y., Esser, R., \& Hu, Y.\ 2003, \apj, 582, 467
\bibitem[Chiu et al.(1998)]{1998SSRv...86..257C} Chiu, M.~C., von-Mehlem, U.~I., Willey, C.~E., et al.\ 1998, \ssr, 86, 257
\bibitem[Ciaravella et al.(2013)]{2013ApJ...766...65C} Ciaravella, A., Webb, D.~F., Giordano, S., \& Raymond, J.~C.\ 2013, \apj, 766, 65
\bibitem[Elliott et al.(2005)]{2005JGRA..110.4103E} Elliott, H.~A., McComas, D.~J., Schwadron, N.~A., et al.\ 2005, Journal of Geophysical Research (Space Physics), 110, A04103
\bibitem[Geiss et al.(1995)]{1995Sci...268.1033G} Geiss, J., Gloeckler, G., von Steiger, R., et al.\ 1995, Science, 268, 1033
\bibitem[Gibson \& Low(1998)]{1998ApJ...493..460G} Gibson, S.~E., \& Low, B.~C.\ 1998, \apj, 493, 460
\bibitem[Gloeckler et al.(1998)]{1998SSRv...86..497G} Gloeckler, G., Cain, J., Ipavich, F.~M., et al.\ 1998, \ssr, 86, 497
\bibitem[Gosling et al.(2007)]{2007JGRA..112.8106G} Gosling, J.~T., Eriksson, S., McComas, D.~J., et al.\ 2007, Journal of Geophysical Research (Space Physics), 112, A08106
\bibitem[Gosling(1990)]{1990GMS....58..343G} Gosling, J.~T.\ 1990, Washington DC American Geophysical Union Geophysical Monograph Series, 58, 343
\bibitem[Gruesbeck et al.(2011)]{2011ApJ...730..103G} Gruesbeck, J.~R., Lepri, S.~T., Zurbuchen, T.~H., et al.\ 2011, \apj, 730, 103
\bibitem[Gruesbeck et al.(2012)]{2012ApJ...760..141G} Gruesbeck, J.~R., Lepri, S.~T., \& Zurbuchen, T.~H.\ 2012, \apj, 760, 141
\bibitem[Howard(2011)]{2011JASTP..73.1242H} Howard, T.~A.\ 2011, Journal of Atmospheric and Solar-Terrestrial Physics, 73, 1242
\bibitem[Hudson, \& Cliver(2001)]{2001JGR...10625199H} Hudson, H.~S., \& Cliver, E.~W.\ 2001, \jgr, 106, 25199
\bibitem[Hu \& Sonnerup(2002)]{2002JGRA..107.1142H} Hu, Q., \& Sonnerup, B.~U. {\"O}.\ 2002, Journal of Geophysical Research (Space Physics), 107, 1142
\bibitem[Kim et al.(2013)]{2013SoPh..284...77K} Kim, R.-S., Gopalswamy, N., Cho, K.-S., et al.\ 2013, \solphys, 284, 77
\bibitem[Klein \& Burlaga(1982)]{1982JGR....87..613K} Klein, L.~W., \& Burlaga, L.~F.\ 1982, \jgr, 87, 613
\bibitem[Ko et al.(1999)]{1999JGR...10417005K} Ko, Y.-K., Gloeckler, G., Cohen, C.~M.~S., et al.\ 1999, \jgr, 104, 17005
\bibitem[Ko et al.(2010)]{2010ApJ...722..625K} Ko, Y.-K., Raymond, J.~C., Vr{\v{s}}nak, B., et al.\ 2010, \apj, 722, 625
\bibitem[Ko et al.(2013)]{2013AIPC.1539..207K} Ko, Y.-K., Raymond, J.~C., Rakowski, C., \& Rouillard, A.\ 2013, Solar Wind 13, 1539, 207
\bibitem[Landi et al.(2012)]{2012ApJ...744..100L} Landi, E., Alexander, R.~L., Gruesbeck, J.~R., et al.\ 2012, \apj, 744, 100
\bibitem[Landi et al.(2012)]{2012ApJ...761...48L} Landi, E., Gruesbeck, J.~R., Lepr, S.~T., Zurbuchen, T.~H., \& Fisk, L.~A. \ 2012, \apj, 13, 761, 1
\bibitem[Lepping et al.(1990)]{1990JGR....9511957L} Lepping, R.~P., Jones, J.~A., \& Burlaga, L.~F.\ 1990, \jgr, 95, 11957
\bibitem[Lepping et al.(2006)]{2006AnGeo..24..215L} Lepping, R.~P., Berdichevsky, D.~B., Wu, C.-C., et al.\ 2006, Annales Geophysicae, 24, 215
\bibitem[Lepping et al.(2011)]{2011SoPh..274..345L} Lepping, R.~P., Wu, C.-C., Berdichevsky, D.~B., et al.\ 2011, \solphys, 274, 345
\bibitem[Lepri, \& Zurbuchen(2004)]{2004JGRA..109.1112L} Lepri, S.~T., \& Zurbuchen, T.~H.\ 2004, Journal of Geophysical Research (Space Physics), 109, A01112
\bibitem[Lepri et al.(2012)]{2012ApJ...760..105L} Lepri, S.~T., Laming, J.~M., Rakowski, C.~E., \& von Steiger, R.\ 2012, \apj, 760, 105
\bibitem[Lindsay et al.(1999)]{1999JGR...10412515L} Lindsay, G.~M., Luhmann, J.~G., Russell, C.~T., \& Gosling, J.~T.\ 1999, \jgr, 104, 12515
\bibitem[Lin et al.(2004)]{2004ApJ...602..422L} Lin, J., Raymond, J.~C., \& van Ballegooijen, A.~A.\ 2004, \apj, 602, 422
\bibitem[Lin \& Forbes(2000)]{2000JGR...105.2375L} Lin, J., \& Forbes, T.~G.\ 2000, \jgr, 105, 2375
\bibitem[Liu et al.(2003)]{2003ApJ...593L.137L} Liu, Y., Jiang, Y., Ji, H., et al.\ 2003, \apjl, 593, L137
\bibitem[Lundquist(1950)]{1950Ark.35...361} Lundquist, S.\ 1950,\ Ark. Fys., Stockholm 2, No.35, 361-365
\bibitem[Lugaz et al.(2005)]{2005ApJ...634..651L} Lugaz, N., Manchester, W.~B., \& Gombosi, T.~I.\ 2005, \apj, 634, 651
\bibitem[Lugaz \& Roussev(2011)]{2011JASTP..73.1187L} Lugaz, N., \& Roussev, I.~I.\ 2011, Journal of Atmospheric and Solar-Terrestrial Physics, 73, 1187
\bibitem[Lynch et al.(2003)]{2003JGRA..108.1239L} Lynch, B.~J., Zurbuchen, T.~H., Fisk, L.~A., et al.\ 2003, Journal of Geophysical Research (Space Physics), 108, 1239
\bibitem[Lynch et al.(2011)]{2011ApJ...740..112L} Lynch, B.~J., Reinard, A.~A., Mulligan, T., et al.\ 2011, \apj, 740, 112
\bibitem[McComas et al.(1998)]{1998SSRv...86..563M} McComas, D.~J., Bame, S.~J., Barker, P., et al.\ 1998, \ssr, 86, 563
\bibitem[Mikic \& Linker(1994)]{1994ApJ...430..898M} Mikic, Z., \& Linker, J.~A.\ 1994, \apj, 430, 898
\bibitem[Miller, \& Vinas(1993)]{1993ApJ...412..386M} Miller, J.~A., \& Vinas, A.~F.\ 1993, \apj, 412, 386
\bibitem[Neugebauer \& Goldstein(1997)]{1997GMS....99..245N} Neugebauer, M., \& Goldstein, R.\ 1997, Washington DC American Geophysical Union Geophysical Monograph Series, 99, 245
\bibitem[Osherovich et al.(1993)]{1993AdSpR..13...57O} Osherovich, V.~A., Farrugia, C.~J., \& Burlaga, L.~F.\ 1993, Advances in Space Research, 13, 57
\bibitem[Ouyang et al.(2015)]{2015ApJ...815...72O} Ouyang, Y., Yang, K., \& Chen, P.~F.\ 2015, \apj, 815, 72
\bibitem[Owens(2018)]{2018SoPh..293..122O} Owens, M.~J.\ 2018, \solphys, 293, 122
\bibitem[Patsourakos et al.(2013)]{2013ApJ...764..125P} Patsourakos, S., Vourlidas, A., \& Stenborg, G.\ 2013, \apj, 764, 125
\bibitem[Qiu \& Yurchyshyn(2005)]{2005ApJ...634L.121Q} Qiu, J., \& Yurchyshyn, V.~B.\ 2005, \apjl, 634, L121
\bibitem[Reinard, \& Biesecker(2009)]{2009ApJ...705..914R} Reinard, A.~A., \& Biesecker, D.~A.\ 2009, \apj, 705, 914
\bibitem[Richardson \& Cane(2004)]{2004JGRA..109.9104R} Richardson, I.~G., \& Cane, H.~V.\ 2004, Journal of Geophysical Research (Space Physics), 109, A09104
\bibitem[Rodriguez et al.(2004)]{2004JGRA..109.1108R} Rodriguez, L., Woch, J., Krupp, N., et al.\ 2004, Journal of Geophysical Research (Space Physics), 109, A01108
\bibitem[Smith et al.(1998)]{1998SSRv...86..613S} Smith, C.~W., L'Heureux, J., Ness, N.~F., et al.\ 1998, \ssr, 86, 613
\bibitem[Song et al.(2014)]{2014ApJ...784...48S} Song, H.~Q., Zhang, J., Cheng, X., et al.\ 2014, \apj, 784, 48
\bibitem[Song et al.(2015a)]{2015ApJ...804L..38S} Song, H.~Q., Chen, Y., Zhang, J., et al.\ 2015, \apjl, 804, L38
\bibitem[Song et al.(2015b)]{2015ApJ...803...96S} Song, H.~Q., Zhang, J., Chen, Y., et al.\ 2015, \apj, 803, 96
\bibitem[Song et al.(2015c)]{2015ApJ...808L..15S} Song, H.~Q., Chen, Y., Zhang, J., et al.\ 2015, \apjl, 808, L15
\bibitem[Song et al.(2016)]{2016ApJS..224...27S} Song, H.~Q., Zhong, Z., Chen, Y., et al.\ 2016, \apjs, 224, 27
\bibitem[Vr{\v{s}}nak et al.(2005)]{2005A&A...435.1149V} Vr{\v{s}}nak, B., Sudar, D., \& Ru{\v{z}}djak, D.\ 2005, \aap, 435, 1149
\bibitem[Wang et al.(2016)]{2016JGRA..121.9316W} Wang, Y., Zhuang, B., Hu, Q., et al.\ 2016, Journal of Geophysical Research (Space Physics), 121, 9316
\bibitem[Wang et al.(2018)]{2018JGRA..123.3238W} Wang, Y., Shen, C., Liu, R., et al.\ 2018, Journal of Geophysical Research (Space Physics), 123, 3238
\bibitem[Wurz et al.(2000)]{2000JGR...10527239W} Wurz, P., Bochsler, P., \& Lee, M.~A.\ 2000, \jgr, 105, 27239
\bibitem[Yan et al.(2018)]{2018ApJ...853L..18Y} Yan, X.~L., Yang, L.~H., Xue, Z.~K., et al.\ 2018, \apjl, 853, L18
\bibitem[Yashiro et al.(2002)]{2002AAS...200.3704Y} Yashiro, S., Gopalswamy, N., Michalek, G., \& Howard, R.~A.\ 2002, Bulletin of the American Astronomical Society, 34, 37.04
\bibitem[Zhang et al.(2012)]{2012NatCo...3..747Z} Zhang, J., Cheng, X., \& Ding, M.-D.\ 2012, Nature Communications, 3, 747
\bibitem[Zhao et al.(2017)]{2017ApJ...845..109Z} Zhao, A., Wang, Y., Liu, J., et al.\ 2017, \apj, 845, 109
\bibitem[Zurbuchen et al.(2003)]{2003AIPC..679..604Z} Zurbuchen, T.~H., Fisk, L.~A., Lepri, S.~T., et al.\ 2003, Solar Wind Ten, 604
\bibitem[Zurbuchen \& Richardson(2006)]{2006SSRv..123...31Z} Zurbuchen, T.~H., \& Richardson, I.~G.\ 2006, \ssr, 123, 31
\bibitem[Zurbuchen et al.(2016)]{2016ApJ...826...10Z} Zurbuchen, T.~H., Weberg, M., von Steiger, R., et al.\ 2016, \apj, 826, 10


\end{thebibliography}

\listofchanges
\end{document}